\documentclass[11pt,a4paper]{article}
\pdfoutput=1
\usepackage[utf8]{inputenc}
\usepackage{amsmath}
\usepackage{amsfonts}
\usepackage{amssymb}
\usepackage{graphicx}
\usepackage[left=2cm,right=2cm,top=2cm,bottom=2cm]{geometry}
\usepackage[bookmarks=true,bookmarksopen=true,colorlinks=true,hypertexnames=true,citecolor=blue,linkcolor=blue,file
color=blue,backref,colorlinks,breaklinks]{hyperref}
\usepackage[affil-it]{authblk}

\usepackage{fancyhdr}
\usepackage{multirow, multicol, makecell}
\usepackage{float}

\usepackage{natbib}
\usepackage{epstopdf}

\title{\textbf{The Impact of Hurricanes on the Oceanographic Conditions in the Exclusive Economic Zone of Cuba}}

\author[1, 2,*]{Dail\'{e} Avila-Alonso}
\author[2]{Jan M. Baetens}
\author[1]{Rolando Cardenas}
\author[2]{Bernard De Baets}

\affil[1]{Laboratory of Planetary Science, Department of Physics, Universidad Central ``Marta Abreu" de Las Villas, 54830, Santa Clara, Villa Clara, Cuba. *davila@uclv.cu}                                                                            

\affil[2]{KERMIT, Department of Data Analysis and Mathematical Modelling, Faculty of Bioscience Engineering, Ghent University, 9000 Ghent, Belgium.}

\date{}

\begin{document}
\maketitle
%

\section*{Abstract}
In this work, we analysed the satellite-based responses of sea surface temperature (SST) and chlorophyll-a (chl-a) concentration in the waters of the Exclusive Economic Zone (EEZ) of Cuba to  hurricanes that crossed the EEZ between 1998 and 2016. We considered two spatial scales to capture the spatially heterogeneous nature of the effects of hurricanes. A first more fine-grained one where we considered 120~km radius disks centered at every consecutive hurricane position within the EEZ (scale~1) and a second more coarse grained one enclosing the entire EEZ (scale~2). We conclude that the hurricanes induced a weak cooling since 75 and 85$\%$ of the SST anomalies at scale~1~and~2, respectively, were smaller than --1$^{\circ}$C. The cooling was mainly caused by the wind, inducing mixing and/or upwelling of subsurface cool waters. The maximum chl-a responses were recorded in the first and second post-storm weeks, with 60$\%$ ranging between --0.01~and~0.04~mg~m$^{-3}$ at scale 1, and between --0.07 and 0.02 mg m$^{-3}$ at scale~2. During those post-storm weeks SST and chl-a anomalies were 18 and 44$\%$ higher at scale 1 than at scale 2, respectively. We argue that the transport of chl-a from the deep chlorophyll maximum and/or the rich coastal waters are the dominant mechanisms determining the post-storm chl-a response in the EEZ. We also found that the magnitude of the Island Mass Effect in the EEZ after the passage of the hurricanes was 89$\%$ higher than before its passage.

\begin{flushleft}
\textbf{Keywords:} Chlorophyll-a concentration, Exclusive Economic Zone of Cuba, hurricanes, remote sensing, sea surface temperature
\end{flushleft}

\section{Introduction}

Tropical cyclones (TCs) have important impacts on the structure and functioning of marine ecosystems, as they incur changes in their physical environment, as well as in their biological assemblages~\citep{fiedler2013}. TCs can induce persistent patches of increased oceanic primary and secondary production \citep{fiedler2013, gauns2016}. \cite{menkes2016}~reported that in some TC-prone regions, the TC contribution to the annual primary production can locally reach up to 20--30$\%$, while nearly 80$\%$ of this TC-induced annual primary production is the result of the biogeochemical response to the 30$\%$ strongest TCs. Climate change may alter the frequency and/or intensity of these phenomena around the world. Future projections suggest that greenhouse warming over the 21$\textsuperscript{st}$ century will lead to fewer TCs, though the frequency of the strongest storms and their associated rainfall amounts will increase~\citep{knutson2010, knutson2013, walsh2016}. In the last years, a rise in Atlantic TC activity has been observed and most studies agree this is a consequence of the increased sea surface temperature (SST) of the Atlantic Ocean~\citep{wang2007}. Thus, we may anticipate that the rise in Atlantic TC activity will continue since models project the greatest warming in the Caribbean Sea during the summer months by the end of the century~\cite [][and references therein]{christensen2013}.

The Atlantic Warm Pool (AWP) is the large body of warm water ($>$ 28.5$^{\circ}$C) extending through the Gulf of Mexico, the Caribbean Sea, and the western tropical North Atlantic during the summer and fall that favours TC formation and intensification~\citep{wang2007}. The waters surrounding Cuba are the first ones that reach temperatures higher than 28.5$^{\circ}$C at the beginning of the Atlantic Hurricane Season (June 1$\textsuperscript{st}$ to November 30$\textsuperscript{th}$) and they remain warmer than 28.5$^{\circ}$C during the entire hurricane season \citep{wang2007, wang2011}. Most North Atlantic TCs typically evolve from tropical waves that regularly move westward off the coast of Africa~\citep{mcadie2009}. After moving into the northern Caribbean or Greater Antilles regions, they usually either move towards the Gulf of Mexico or recurve and accelerate in the North Atlantic. Some TCs can recurve after reaching the Gulf of Mexico, while others continue westward to make landfall in the United States of America or Mexico~\citep{bowditch2002}. In any case, and according to the climatological hurricane trajectories in the North Atlantic, the waters of the Exclusive Economic Zone (EEZ) of Cuba can be affected by TCs during the entire hurricane season (\textit{http://www.nhc.noaa.gov/climo/}). All this makes the EEZ of Cuba a hot-spot area for the study of TC effects on marine ecosystems. 

Considering the maximum sustained wind speed (i.e., the highest one-minute average wind at an elevation of 10~m with an unobstructed exposure at a particular point in time, \textit{http://www.nhc.noaa.gov/
aboutgloss.shtml}), TCs can be classified into the following categories: tropical depression, tropical storm and hurricane, depending on whether sustained winds are, respectively, lower than 18~m~s$^{-1}$, between 18 and 32~m~s$^{-1}$, or equal and larger than 33~m~s$^{-1}$~\citep{mcadie2009}. During a typical Atlantic Hurricane Season, 10 or 11 tropical storms form, of which around 6 evolve into hurricanes~\citep{willoughby2007}. For example, of the 521~TCs that crossed or passed the United States of America coastline in the period 1851--2006, 278~were classified as hurricanes~\citep{mcadie2009}. In the same way, of the 38~TCs that have crossed the EEZ of Cuba in 1998--2016, 18~were hurricanes, 11~tropical storms and 9~tropical depressions according to the National Hurricane Center (NHC) (\textit{http://www.nhc.noaa.gov/data/tcr/}). So, hurricanes are common meteorological phenomena in this TC-prone area. Besides, hurricanes and in particular the strongest ones induce the highest oceanographic variability given that the ocean response depends on the TC strength \mbox{\citep{hanshaw2008, menkes2016}}.

The EEZ of a country is a maritime area where it exercises rights for exploring, \mbox{exploiting,} conserving, and managing the natural marine resources. Countries must manage \mbox{fisheries} within their EEZ in order to derive indicators for marine ecosystems at the national level \mbox{\citep{prescottAllen2001}}. Fisheries are ultimately linked to the patterns of phytoplankton production. This explains why several studies assessed the influence of oceanographic variables in general, and primary production (or chlorophyll-a (chl-a) concentration) of phytoplankton in particular, on \mbox{fisheries} at the EEZ level \mbox{\cite[e.g.,][]{dambacher2012, mustapha2013, deleo2014}}. Besides, other studies considered the effect of large-scale environmental drivers such as global warming and El Ni\~{n}o Southern Oscillation on oceanography and fishing in EEZs \citep[e.g.,][]{murphy2001, domokos2009, barange2014, watson2015}. Extreme environmental disturbances caused by TCs lead to increased nutrient concentrations in the upper ocean layer, which stimulates phytoplankton growth and hence can increase fish biomass production~\mbox{\citep{hung2014}}. For instance, increased zooplankton biomass and feeding success of planktivorous fish following an increase in phytoplankton abundance induced by the passage of a TC in the Pacific Ocean has been reported \mbox{\citep{fiedler2013}}. Despite the anticipated future increase in the frequency of the strongest storms and the need to understand the phytoplankton response to changing environmental conditions, there are no studies on the oceanographic variability due to the passage of hurricanes in the waters of the EEZ of Cuba.

Satellite remote sensing is suitable to assess the oceanographic conditions at large spatial scales such as EEZs, since it provides a large amount of data with both a high spatial and temporal resolution \mbox{\citep{murphy2001, pieralice2014, cristina2015}}. Moreover, satellite data are promising sources of information to further our understanding of the TC-induced oceanographic variability~\mbox{\citep{son2007}} given that \textit{in situ} measurements under extreme meteorological conditions are difficult to obtain. For these reasons, we assess the oceanographic variability of the waters of the EEZ of Cuba to the passage of hurricanes using remote sensing data as primary source of information. We first present a brief description of the waters of the EEZ of Cuba and the methods used, including the data sets containing the relevant physical and biological variables. Next, we address the mean daily evolution of anomalies (averaging all hurricane responses) of important explanatory and response variables. We then provide examples of the spatial chl-a distribution in the EEZ and regions with frequent chl-a increases after the passage of hurricanes in the Gulf of Mexico. We also report on the Island Mass Effect (IME) that is induced after the passage of hurricanes in the ocean waters of the Cuban EEZ. Then, regression analyses between the response variables after the passage of hurricanes and explanatory variables are presented. Finally, we present our analysis of the results and a discussion of possible explanations for the observed hurricane-induced SST decrease and surface chl-a increase.

\section{Materials and Methods}
\subsection{Study area and hurricanes}
The waters of the EEZ of Cuba encompass those of the Caribbean Sea and the Gulf of Mexico Large Marine Ecosystems (LMEs) (Figure~1A). Given that those LMEs show different biophysical traits, such as their bathymetry, hydrography, productivity and trophically dependent populations, the waters of the EEZ of Cuba are spatially heterogeneous. The shallow, productive and optically complex marine waters surrounding Cuba are limited to four specific regions, as a consequence of the extent of its marine shelf. The latter drops to 400 m depth or more near its edge, resulting in very shallow coastal areas and very deep ocean waters in the EEZ (Figure~1B). Besides, near the edge, lines of fringing keys and reefs can be found, which have led to the formation of four shallow macrolagoons between the coastline and the outer border of the marine shelf, i.e., Los Colorados and Sabana-Camag\"{u}ey Archipelagos in the northwestern and northcentral part of the Cuban shelf, respectively, the Gulfs of Ana Maria and Guacanayabo covering the southeastern part and the Gulf of Batabano in the southwestern part of the Cuban shelf \citep{claro2001}  (Figure~1C). 

The marine waters of the EEZ of Cuba are oligotrophic~\citep{leal2001, lugioyo2007}, just as the oceanic regions of the Gulf of Mexico and the Caribbean Sea LMEs \citep{okolodkov2003, callejas2012, gomez2014, aguirre2015}, where the reflectance spectra show a maximum at the blue waveband~\citep{gomez2014}. These could be considered as Case~1 waters according to the clasification of~\cite{morel1980} under normal meteorological conditions, since their optical properties are dominated by phytoplankton and associated degradation products. The shapefile of the EEZ of Cuba was obtained from the world EEZ product (version~9) (\textit{http://www.marineregions.org/downloads.php}) of the Flanders Marine Institute, where the inner boundary of the EEZ (i.e., the baseline) in Figure~1C is used as a proxy for the low-water line.

We assessed the oceanographic variability induced by hurricanes that crossed the EEZ of Cuba over the period 1998--2016 in agreement with one of the selection criteria employed by \mbox{\cite{deScally2008}} to study the TC activity across the Cook Islands (Southwest Pacific). Specifically, we included in our study TCs that at some point of their trajectory within the EEZ reached the hurricane category, i.e., 18 hurricanes according to the NHC reports (\mbox{\textit{http://www.nhc.noaa.gov/data/tcr/}}). The starting year of our study period is motivated by the fact that 1998 is the first year for which satellite-derived chl-a concentration data from the Sea-viewing Wide Field-of-view Sensor (SeaWiFS) are available for the entire Atlantic Hurricane Season~\citep{babin2004}. We obtained the hurricane trajectories from the International Best Track Archive for Climate Stewardship (IBTrACS v03r03) \citep{knapp2010} (\textit{ftp://eclipse.ncdc.noaa.gov/pub/ibtracs/v03r09/all/shp/}).

\subsection{Spatial and temporal scales}
In previous studies of other regions, researchers considered different spatial scales to assess the TC-induced oceanographic variability. For example, \cite{babin2004} constructed cells of 0.5$^{\circ}$~latitude north and south and 1$^{\circ}$ longitude west and 2$^{\circ}$ longitude east for each hurricane track point analysed, while other authors have used 200 km radius disks centered at the consecutive TC positions~\citep{menkes2016, shropshire2016} and square areas of different sizes located at variable distances from the TC trajectory~\citep{wang2008, merritt2013}. On the other hand, \cite{son2007} and \cite{farfan2014} analysed a band of 700 km and 15 km wide, respectively, centered on storm tracks, while~\cite{foltz2015} and~\cite{Parkeretal2017} assessed TC activity across a large region of the Western Subtropical North Atlantic and the Great Barrier Reef of Australia, respectively.

Given the spatial structure of hurricanes, their oceanic effects are expected to be spatially heterogeneous. Thus, we used two scales in order to capture the spatially heterogeneous nature of the TC effects: One limited to the trajectory of the hurricanes in the EEZ (scale~1) and another one considering the response of the entire EEZ (scale~2). We assessed the oceanographic variability along the hurricane trajectory in 120 km radius disks centered at every consecutive 6-hourly hurricane position (\textit{http://www.aoml.noaa.gov/hrd/hurdat/hurdat2.html}) within the EEZ. This radius accounts for the mean eyewall width~\cite [both concentric and single eyewalls, see Table~1 in] [] {hence2012} of category~4~and~5~hurricanes in the Atlantic and Northwest Pacific basins between 1998 and 2007~\citep{hence2012}. So, scale~1 accounts for the effects of the heaviest rainfalls and strongest surface winds associated to the eyewall of the hurricanes. On the other hand, an assessment for the entire EEZ response is possible since it was completely exposed to hurricane effects. Indeed, the outer spiral rainbands of almost all analysed hurricanes covered the entire EEZ, as was confirmed by a visual inspection of the GOES-13 visible satellite images (\textit{http://www.goes.noaa.gov/browsh.html}). Therefore, scale~2 accounts for the effects of both the eyewall and the outer spiral rainbands of hurricanes. 

We analysed the environmental conditions before and after the passage of the hurricanes at a daily resolution, which is the highest temporal resolution available for gridded ocean color data~\citep{menkes2016}. This also allowed us to assess the oceanographic conditions during the pre-storm week (i.e., days --10 to --3 before hurricane passage) as a benchmark for comparison with the four post-storm weeks (i.e., from day 0 to +30, where day 0 refers to the day the hurricane entered the study area)~\citep{menkes2016}. Daily and weekly arithmetic means were computed for each analysed variable at both scales. We calculated the mean daily value of the pixel data within the disks of 120~km radius along the hurricane trajectories (scale~1) or the entire EEZ (scale~2). For the former scale, the obtained mean values of the consecutive disks were averaged to retrieve the daily mean along the entire hurricane trajectory, in agreement with the approach followed by~\mbox{\cite{babin2004}}, \mbox{\cite{lloyd2011}} and \mbox{\cite{lloyd2011a}}. Thus, observations at this scale may be \mbox{considered} as Lagrangian measurements \mbox{\citep{lloyd2011, lloyd2011a}}, while observations at scale~2 are Eulerian measurements. Weekly means were then calculated from these daily means. We also computed standardized anomalies by subtracting the mean pre-storm week value from the daily values (from day --10 to day 30) for every hurricane and then calculating the mean daily anomalies across all hurricanes, in agreement with the procedure followed by~\mbox{\cite{menkes2016}}.

\subsection{Data sets}
The response of the upper ocean to a particular TC depends on several atmospheric and oceanic variables~\cite [][and references therein]{babin2004}. Intense winds associated with TCs cause vigorous mixing of the upper ocean and upwelling, which cools the upper layers~\citep{emanuel2001}, allowing the transport of nutrients from below the nitracline and/or chl-a from the deep ocean to the surface, which increases the chl-a concentration after the passage of TCs. Besides, heavy rainfall can also cool the surface waters and give rise to an increase of the chl-a concentration~\citep{lin2016}. For these reasons, we considered the SST and chl-a concentration (as a proxy for the abundance of phytoplankton) as the primary variables accounting for the physical and biological responses to hurricanes, respectively, and the wind speed, rainfall amount and vertical distribution of nutrients as possible environmental factors influencing the magnitude of the post-storm responses. We also considered the transit speed of the hurricanes in the EEZ as this influences the amplitude of the TC energy input into the ocean~\citep{vincent2012a} and largely controls the TC-induced SST cooling~\citep{mei2012}.

\subsubsection{Satellite-derived data}
Assessing the response to hurricanes through satellite observations is an obvious albeit difficult task, as responses are often masked or biased by the extreme environmental conditions intrinsic to these meteorological phenomena~\citep{menkes2016}. Yet, since we are studying a relatively small region, a high spatial resolution and data coverage allows us to better capture environmental variability, enabling a better understanding of the mentioned responses. Thus, we have taken this into account when selecting data.

\paragraph{Infrared/Microwave radiometers data}
The primary wind speed data used to analyse TCs are derived from imagery of geostationary and low-Earth-orbiting satellites, remotely sensed data collected by reconnaissance aircraft and \textit{in situ} data~\cite [reviewed by] [] {stiles2014}. The former have a low accuracy under high-wind and rainy conditions and in general underestimate wind speed during extreme meteorological events~\citep{carswell2000, donelan2004, brennan2009}. However, satellite measurements of winds close to the ocean surface can provide information over extensive areas of oceans~\citep{stiles2014}, which is necessary to compare the oceanographic response to hurricanes across the considered scales. Thus, we used the Cross-Calibrated Multi-Platform (CCMP, version~2.0) 6-hourly gridded ocean vector wind (to a height of 10~m) data~\citep{atlas2011}, produced by Remote Sensing Systems (data available at \textit{ftp://ftp.remss.com/ccmp/v02.0/}), which have been used in other studies assessing TC effects  \mbox{\cite[e.g.,][]{fu2014, peng2015, lin2016}.} CCMP provides a consistent, gap-free long-term time series of ocean surface wind vector fields, as it combines cross-calibrated satellite microwave winds, buoy and model wind data \citep{atlas2011} to arrive at gridded Level 3 products with a high spatial resolution (0.25$^{\circ}$ $\times$ 0.25$^{\circ}$). We determined the daily wind speed by averaging the 6-hourly products. 

We used the Tropical Rainfall Measuring Mission (TRMM) Multi-satellite Precipitation Analysis (TMPA), which takes advantage of the increasing number of satellite-borne precipitation-related sensors to produce high quality microwave/infrared daily accumulated precipitation (DAP) estimates at improved spatial resolution (0.25$^{\circ}$ $\times$ 0.25$^{\circ}$)~\citep{huffman2007, huffman2010}. We used the post-real-time data~\cite [version~7,] [] {huffman2013}, which employs additional data sources, like the TRMM Combined Instrument estimates. The latter combines data from both TRMM Microwave Imager (TMI) and TRMM Precipitation Radar (PR)~\citep{huffman2010}. PR is especially useful for identifying the regions where the TMI is significantly underestimating rain rates, for example in the eyewall of the hurricanes~\citep{zagrodnik2013}.

We derived SST data using the Operational SST and Sea Ice Analysis (OSTIA) Reprocessed (1998--2007) and Near Real Time (2008--2016) Level 4 SST products~\citep{donlon2012} provided by the Copernicus Marine Environment Monitoring Service (\textit{http://marine.copernicus.eu}). OSTIA merges both infrared and microwave radiometer data, together with \textit{in situ} observations. The analysis is performed at a spatial resolution of 0.05$^{\circ}$ $\times$ 0.05$^{\circ}$. Infrared radiometers are strongly influenced by water vapour and cloud contamination, but they have a higher spatial resolution than microwave radiometers~\citep{wentz2000}. In contrast, the latter are insensitive to cloud interference since microwaves penetrate clouds easily, but raindrops can significantly attenuate and scatter microwaves~\citep{wentz2000}. In order to have data with a high spatial resolution and coverage, and taking into account that both cloud coverage and heavy precipitation are inherent to hurricanes~\citep{mcadie2009}, we argue that merged infrared and microwave satellite SST data can be more accurate, since they combine the advantages of both data sets~\cite [] [and references therein]  {devi2015}.

\paragraph{Infrared radiometer data}
    
The chl-a images were obtained from the multisatellite merged global data of the GlobColour project 
(\textit{http://globcolour.info}), developed, validated, and distributed by ACRI-ST, France. These data result from merging multiple mission observations into a single product with a better spatial coverage than for individual missions~\citep{maritorena2005, maritorena2010}. After TC passage,  marine waters could basically be Case~2 of the~\cite{morel1980} classification (i.e., waters with high concentrations of coloured dissolved organic matter (CDOM) and total suspended matter). \cite{acker2009} reported the advection of coastal waters containing high concentrations of CDOM into the southeastern Gulf of Mexico after the passage of hurricane Katrina (2005) near the coast of Florida. Thus, we selected the CHL-OC5 data (computed with the OC5~algorithm) at a spatial resolution of 0.0417$^{\circ}$ $\times$ 0.0417$^{\circ}$, which is the highest one within the framework of the GlobColour project. The method used to determine the CHL-OC5 data is empirical, and derived from the OC4/SeaWiFS algorithm of NASA (or OC3M-547 for Moderate Resolution Imaging Spectroradiometer (MODIS) and OC4E for Medium Resolution Imaging Spectrometer (MERIS))~\citep{gohin2011}. It uses the 412 and 555~wavelenghts accounting for the effects of CDOM and suspended matter, respectively~\citep{gohin2002, gohin2011}. \cite{loisel2017}~compared and validated several chl-a algorithms for Vietnamese coastal waters, and reported that the OC5 algorithm showed the best and most stable results, which is coherent with previous studies stressing the potential of OC5 for the assessment of chl-a across a large range of optically complex waters. 

Besides, the CHL-OC5 data have a better spatial coverage than the other chl-a data of the GlobColour project. For instance, on average 32$\%$ of pixels of the CHL1~GSM images (see \textit{http://www.
globcolour.info/CDR$\textunderscore$Docs/GlobCOLOUR$\textunderscore$PUG.pdf} for details) lie in the EEZ of Cuba (for hurricanes and days analysed), while this amounts to 56$\%$ for CHL-OC5 images. We also used the data of the coloured dissolved and detrital organic material absorption coefficient at 443~nm (CDM) computed with the Garver, Siegel and Maritorena model~\cite [GSM,] [] {maritorena2010} and the phytoplankton absorption coefficient at 443~nm (a$_{ph}$), provided by MODIS Aqua (\textit{https://oceancolor.gsfc.
nasa.gov}) at a spatial resolution of 0.0417$^{\circ}$ $\times$ 0.0417$^{\circ}$.

\paragraph{Altimeter data}
Given the geographical location of the EEZ of Cuba and the complex system of marine currents in the Gulf of Mexico and the Caribbean Sea LMEs, which likely influence the distribution of chl-a, we used the Salto/DUACS gridded multimission altimeter products processed and distributed  by the Copernicus Marine Environment Monitoring Service (\textit{http://marine.copernicus.eu}). The zonal and meridional components of the absolute geostrophic currents were obtained from the Map of Absolute Dynamic Topography (0.25$^{\circ}$ $\times$ 0.25$^{\circ}$ of spatial resolution) and used to determine the vector field of the geostrophic marine currents. 

\subsubsection{World Ocean Data}
The vertical distribution of nutrients and phytoplankton govern the biogeochemical response after the TC passage~\citep{babin2004}. Nitrogen availability tends to limit phytoplankton productivity throughout much of the surface of the low-latitude oceans, including the Gulf of Mexico and the Caribbean Sea LMEs~\cite [see Figure 3 in] [] {moore2013}. Considering that the upward transport of nitrate depends not only on the mixing, but also on the depth from which nitrate originates~\citep{omand2015}, we used the nitrate profiles data of the World Ocean Atlas 2013, version~2 (\textit{https://www.nodc.noaa.gov/OC5/woa13/}), to estimate the climatological depth of the nitracline (DNO$_{3}$) and considered this feature as an important biogeochemical factor constraining the biological response after the TC passage \citep{menkes2016}. The used product is a set of objectively analysed (1$^{\circ}$ $\times$ 1$^{\circ}$) climatological fields of \textit{in situ} nitrate concentration at standard depths for monthly composite periods (\textit{https://www.nodc.noaa.gov/OC5/woa13/}). We estimated DNO$_{3}$ following the procedure in \cite{menkes2016}, i.e., by identifying the depth where the concentration of NO$_{3}$ becomes higher than~1 $\mu$M and DNO$_{3}$ is set to zero if surface NO$_{3}$ concentration exceeds this threshold. The depth of the deep chlorophyll maximum (DCM) is also an important factor contributing to the chl-a increase after the TC passage~\citep{menkes2016}, but we did not include it in our analysis since only a limited number of chl-a profiles from the World Ocean Atlas Database were available for the EEZ of Cuba and also they only extended to depths of 100 m.

\subsubsection{Data derived from the National Hurricane Center}
NHC best track reported times and positions were used (\textit{http://www.aoml.noaa.gov/hrd/hurdat/
hurdat2.html}) to calculate the transit speed of the hurricanes within the EEZ of Cuba, according to the procedure outlined by~\cite{babin2004} and~\cite{gierach2008}. Transit time was the time difference between the first and last TC track point within the EEZ of Cuba, while the transit distance was determined by summing the distances between the consecutive track points within the EEZ. For the hurricanes Ida (2009) and Tomas (2010), there is only one track point inside the EEZ (Figure~2), so we determined their transit speed using the points located just beyond the border of the EEZ. On the other hand, we used the NOAA's Historical Hurricane Tracks online tool (\textit{https://coast.noaa.gov/hurricanes}) to assign the Saffir-Simpson category to hurricanes within the EEZ on the basis of their maximum sustained wind speed reported by the NHC (\textit{http://www.aoml.noaa.gov/hrd/hurdat/hurdat2.html}).

\subsection{Statistical analysis}
In order to identify the main environmental factors (i.e., explanatory variables or predictors) governing the SST and chl-a concentration (response variables) at both scales after a hurricane passage, we considered the wind speed, DAP and transit speed as possible explanatory variables of the SST response, and those explanatory variables together with the SST and DNO$_{3}$ as possible explanatory variables of the chl-a concentration response. We found that predictors of the chl-a concentration showed multicollinearity (variance inflation factor (VIF) ranging from 1.34 to 2.99 along the trajectory, and from 1.29 to 2.34 for the entire EEZ), and similarly for the SST predictors. This indicates an almost linear relationship among some of the predictors~\cite [reviewed by] [] {mason1991}. So, it is not straightforward to determine the effects of the explanatory variables independently. Taking into account that mean VIF values considerably larger than one can indicate serious multicollinearity problems influencing the least squares estimates and that multicollinearity does not affect the overall fit of the model~\citep{kutner2005}, we performed two statistical analyses: (1) single-factor linear regressions to explore the relationships between the response variables and individual predictors and (2) multiple linear regressions to estimate the extent to which the studied environmental factors can explain the variance in SST and chl-a concentration.

\section{Results}
Table 1 lists the hurricanes that crossed the EEZ of Cuba between 1998 and 2016 and provides a summary of relevant physical characteristics. In general, 13 hurricanes made landfall in Cuba (Figure~2), while 12 were classified as major hurricanes (i.e., $\geqslant$ category 3 on the Saffir-Simpson Hurricane Scale (SSHS)) at some point along their trajectory inside the EEZ (Table~1). Hurricanes showed diverse intensification patterns (i.e., changes of the SSHS category along their trajectories, see Table~1). For instance, some hurricanes got stronger, such as Lili (2002), Charley (2004), Rita (2005) and Gustav (2008), while others weakened, such as Michelle (2001), Ivan (2004) and Ike (2008). Finally, the SSHS of other hurricanes, such as Katrina (2005) and Wilma (2005) did not change as they progressed. Essentially, the SSHS of 50$\%$ of the hurricanes increased after entering the EEZ at some point of their trajectory. On the other hand, Paloma (2008) and Paula (2010) weakened from hurricanes to tropical depressions (Figure~2 and Table~1). For the latter, we only considered the points along their trajectory where they were classified as hurricanes in order to only account for the oceanographic variability induced by hurricane strength, which can be different from that induced by TCs of lower category. For instance, it has been reported that hurricanes can lead to a mean SST cooling of approximately 1.44$^{\circ}$C in an area of 5$^{\circ}$ $\times$ 5$^{\circ}$ centered at the TC position for translation speeds from 1 to 9 m s$^{-1}$, while tropical depressions lead to a mean SST cooling of approximately 0.46$^{\circ}$C~\cite [see Figure 1c in] [] {mei2012}.

\begin{table}[H]
\centering
\caption{Characteristics of the hurricanes studied. For the Saffir-Simpson Hurricane Scale (SSHS), different numbers indicate SSHC changes along the trajectory in the Exclusive Economic Zone (EEZ) of Cuba, the first and the last numbers indicate the category when entering and leaving the EEZ, $\mathrm{<1}$ indicates that the hurricane weakened and became a TC of lower intensity.} 
\begin{tabular}{p{1.2cm}p{1.8cm}p{2.2cm}p{1.8cm}p{1.5cm}p{1.8cm}}
\hline 

\multirow{3}{*}{\textbf{Season}} & \multirow{3}{*}{\textbf{Hurricane}} & \multirow{3}{*}{\textbf{SSHS}} & \textbf{Transit distance (km)} & \textbf{Transit time (h)} & \textbf{Transit speed (km h$^{-1}$)} \\

\hline 
1998 & Georges & 1--2 & 868.6 & 42 & 20.68 \\ 
 
2001 & Michelle & 4--3--1 & 624.9 & 42 & 14.9\\ 
 
2002 & Isidore & 1--2--1--2--3 & 498.6 & 42 & 11.87 \\ 
 
     & Lili & 1--2 & 557.5 & 24 & 23.2 \\ 

2004 & Charley & 2--3 & 298.1 & 12 & 24.8 \\ 
 
     & Ivan & 5--4 & 637.9 & 36 & 17.7 \\ 
 
2005 & Dennis & 4--3--4--3--1 & 719.0 & 30 & 24.0 \\ 

     & Katrina & 3 & 161.4 & 12 & 13.5 \\ 

     & Rita & 3--4--5 & 176.5 & 12 & 14.7 \\ 

     & Wilma & 2 & 255.9 & 12 & 21.33 \\ 
 
2008 & Gustav & 3--4--3 & 433.4 & 18 & 24.08 \\ 

     & Ike & 3--4--3--2--1--2 & 1259.3 & 72 & 17.5 \\ 

     & Paloma & 4--2--$\mathrm{<1}$ & 145.8 & 7 & 20.8\\ 

2009 & Ida & 1 & 205.0 & 12 & 17.1\\ 
 
2010 & Paula & 1--$\mathrm{<1}$ & 109.5 & 12 & 9.1 \\ 

     & Tomas & 1 & 254.9 & 12 & 21.2 \\ 

2012 & Sandy & 2--3--2 & 105 & 3.35 & 31.3 \\ 

2016 & Matthew & 4--3 & 155.8 & 12 & 13.0 \\ 
\hline 
\end{tabular} 
\end{table} 

\subsection{Mean daily responses}
Figure~3 displays the mean daily anomalies of wind speed, DAP, SST and chl-a concentration at scales~1~and~2 for all hurricanes. The maximum wind speed anomalies occurred during days --2 to +2, irrespective of the spatial scale (Figure~3A), though they were 19$\%$ higher at scale~1 compared to scale~2. The highest wind speed anomaly was observed at day +1. The drop of the wind speed on day~0 was caused by the light winds in the eye of major hurricanes. Still, the wind speed anomalies that day were high at both scales due to the fact that they represent a spatial mean over a 120~km radius disk centered at the TC position and throughout the entire EEZ. Thus, we account for the eyewall, where the highest wind speed is recorded. After day +2 the wind speed anomalies dropped to pre-storm levels or even lower, indicating that the hurricane moved away from the study area. A second rise occurred at the beginning of the fourth post-storm week (Figure~3A), as a consequence of the overlap between this period for a given hurricane and the pre-storm weeks and days --2 and --1 of a second hurricane of the season, as well as the effects of other meteorological phenomena.

Mean DAP anomalies showed the most conspicuous temporal variability during the analysed period (Figure~3B). Unlike the wind speed anomalies, DAP anomalies were high in the pre-storm week, irrespective of the spatial scale, and at days --2 and --1 (at scale~1). 44 and 66$\%$ of the studied hurricanes showed a higher DAP in the pre-storm week than in the first post-storm week, at scale~1~and~2, respectively. At day +1 DAP anomalies differ between the scales, with the anomalies along the trajectory being 54$\%$ higher than those in the EEZ. A considerable drop of the DAP anomalies occurred between days +2 to +4 at both scales, after which a clear pattern cannot longer be discerned. In general, the maximum positive anomalies occurred in the first, third and fourth post-storm weeks, with the highest values in the third post-storm week (Figure~3B). 

The mean SST and chl-a anomalies do not differ much during the pre-storm week for both scales (Figures~3C~and~3D), though the chl-a anomaly peaks at day --5 at scale~2 (Figure~3D). More pronounced differences between the scales were observed in the first and second post-storm weeks, with SST and chl-a anomalies being 18 and 44$\%$ higher at scale~1 than at scale~2, respectively (Figures~3C~and~3D). It can be observed in Figure~3C that cooling started at day~0, and continued up to the last post-storm week, where the most negative mean SST anomaly (i.e., -0.9$^{\circ}$C) occurred (Figure~3C). In general, 75 and 85$\%$ of the SST anomalies for the individual hurricanes at scale~1~and~2, respectively, were smaller than --1$^{\circ}$C (Figures~4A~and~4B). The largest variability of anomalies was observed at scale~1 ranging from --2.28 to 0.45$^{\circ}$C (Figures 4A~and~4B), though differences between the scales are not as clear as for the chl-a concentration data as we will see below. 

The mean chl-a anomalies were positive during the entire post-storm period, indicating a chl-a concentration increase after the passage of the hurricanes (Figure~3D). The highest ch-a anomalies at scale~1 occurred during the first and second post-storm weeks, with a first peak from day~+2 to +4, and a second one at day~+13. During the third post-storm week, a decrease of the chl-a anomalies is observed, which is followed by a chl-a rise in the fourth post-storm week (Figure~3D). In contrast, the mean post-storm chl-a anomalies at scale~2 showed a consistent response, with the highest increase in the fourth post-storm week. The daily chl-a anomalies associated with individual hurricanes during the first and the second post-storm weeks are displayed in Figures~4C~and~4D. From these figures, we observe a large proportion of weak hurricane chl-a responses, where only 2$\%$ of induced chl-a anomalies exceed 0.1 mg m$^{-3}$ at scale~1. At this scale, 60$\%$ of the chl-a anomalies lie between --0.01 and 0.04 mg m$^{-3}$, while they lie between --0.07 and 0.02 mg m$^{-3}$ at scale~2. Besides, anomalies are highly variable at scale~1 (from --0.04 to 0.18 mg m$^{-3}$), leading to a more positive bias than at scale~2 (Figures~4C~and~4D).

\subsection{Spatial chl-a distribution}
We identified spatial patterns in the chl-a concentration in the EEZ after the passage of hurricanes that agree with the ones reported by \cite{perez2014}, i.e., regional and global \mbox{chl-a} increments throughout the entire EEZ (Figure~5). However, examples of the latter were not manifold due to cloud-induced gaps and incomplete spatial coverage as a consequence of \mbox{satellite} orbits. Regional increments involve local chl-a rises of up to 3.25~mg~m$^{-3}$, such as after the passage of Isidore/Lili (2002) at the San Antonio Cape and the bloom observed after the passage of Michelle~(2001) \mbox{(Figures~5C~and~5D)}. The latter also gave rise to a global increment in the EEZ \mbox{(Figure~5D)}. In both cases (i.e., local and global chl-a increments), the chl-a quantities and distributions differ significantly from the ones during the pre-storm period (Figures~5A~and~5B). 

Local increases were mainly observed close to the inner boundary of the EEZ and showed a peculiar pattern, which suggests that they could have originated from the transport of chl-a rich coastal waters to the deep waters of the EEZ. \cite{perez2014} reported the formation of chl-a filaments extending from the western coast of Cuba to the Gulf of Mexico and the Yucatan Basin after the passage of cold fronts, easterly tropical waves, low-pressure systems and hurricanes Ivan~(2004) and Wilma~(2005). They argued that those filaments were possibly caused by advection of coastal waters. So, the transported waters should present both a high chl-a concentration and also a high CDOM concentration. \cite{acker2009} reported an elevated a$_{ph}$ together with CDOM and detrital absorption coefficient in waters being transported from the West Florida Shelf to the Tortugas Gyre circulation, where a phytoplankton bloom was recorded after the passage of Katrina~(2005) \citep{shi2007}. This bloom was attributed to an augmented nutrient supply caused by wind-driven upwelling and vertical mixing \citep{shi2007}. However, the findings of \cite{acker2009} indicate that some highly absorbing coastal waters transported off the shelf by Katrina were entrained in the gyre circulation, which also contributed to the augmented chl-a concentration during the bloom. 

Figure~6 displays an example of a local chl-a increase after the passage of Isidore/Lili (2002), as well as the corresponding a$_{ph}$ and CDM maps. We selected this example since it shows a very peculiar chl-a distribution. We observe a good agreement between the chl-a distribution and the vector field of the geostrophic marine currents (Figure~6A), suggesting that the latter governs the spatial chl-a pattern. Besides, we see a perfect match between the pattern of the chl-a and those of a$_{ph}$ and CDM (Figures~6A,~6B~and~6C), which suggest that those local chl-a rises were a consequence of coastal optically complex waters transported by the marine currents. On the other hand, Figure~6D shows an SST map with an overlay of chl-a contours (0.13~mg~m$^{-3}$), indicating that the chl-a filaments extend over warm surface waters. So, the absence of low temperatures supports the suggestion that those filaments are not the result of local upwelling. This is in agreement with the conclusions of~\cite{jones2015}, who performed a similar analysis for a high-chl-a plume extending from the continental shelf of the northern Gulf of Mexico to the central basin.

We also identified three regions in the Gulf of Mexico, specifically along the Loop Current, where a chl-a rise typically occurs after the passage of hurricanes (Figure~7). The chl-a rise in region~I (eastern Campeche Bank) was observed after the passage of Isidore/Lili (2002), Ivan (2004) (Figure~7A), Gustav/Ike (2008) (Figure 7C) and Paula (2010). The chl-a rise in region~II (central Gulf of Mexico) was observed after Michelle (2001), Isidore/Lili (2002) (Figure~5C), Ivan (2004) (Figure~7A), Wilma (2005), Gustav/Ike (2008) (Figure~7C) and Paula (2010). Finally, the chl-a rise in region~III (southwest of the Florida Peninsula) was observed after the passage of Katrina (Figure~7B) and Rita (2005), Gustav, Ike (Figure~7C) and Paloma (2008). In all cases, chl-a increases lasted approximately between one and two weeks. These blooms share the common feature that they occurred in regions with cyclonic eddies, for example Campeche Bank eddy and Tortugas Gyre in regions I and III, respectively. Besides, the formation of a large cyclonic eddy occurs around region~II as a consequence of the retraction of the Loop Current~\citep{zeng2015}. Chl-a increments observed in region~II correspond to hurricanes with post-storm weeks in the period of the strongest Loop Current retraction (i.e., from September to December~\citep{zeng2015}). In general, these cyclonic eddies intensify after the passage of hurricanes, inducing a strong upwelling and consequent increase of the chl-a concentration~\citep{walker2005, gierach2008, acker2009}. So, those regions in the Gulf of Mexico may be considered as blooming-prone and thereby positively impact the chl-a concentration in the EEZ of Cuba after the passage of hurricanes as a consequence of the horizontal transport of chl-a. 

\subsection{Island Mass Effect}
The IME in Cuban marine waters is recorded for the first time in this study. Moreover, is the first time that the IME is quantified after the passage of hurricanes globally. The IME refers to an increase of primary production of phytoplankton (or chl-a concentration) in marine waters surrounding islands. It was reported for the first time in 1956 in Hawaii~\citep{doty1956}, while the environmental drivers governing its magnitude were determined only recently~\citep{gove2016}. Accordingly, and given the characteristics of the Cuban marine shelf (e.g., no gradual bathymetric slope, Figure~1B)~\citep{claro2001}, its magnitude in the waters of the EEZ is expected to be low or negligible under normal meteorological conditions. However, it could play a role under extreme meteorological conditions. In this study, we quantified the magnitude of the IME in the waters surrounding Cuba before and after the passage of hurricanes. We analysed three locations with a significant ($\mathrm{\textit{p}}<0.05$) post-storm chl-a concentration increase along transects from the baseline of the EEZ (the inner boundary of the EEZ) to the open ocean (beyond the outer boundary of the EEZ) (see Table~2 for hurricanes, regions and dates analysed). 

\begin{table}[htb]
\centering
\caption{Hurricanes, dates (before and after the passage of hurricanes across the Exclusive Economic Zone (EEZ)) and locations selected to study the Island Mass Effect (IME): coordinates of the transects (from the baseline of the EEZ of Cuba to the open ocean) and locations. Red, yellow, and blue transects were used to quantify the magnitude of the IME for hurricanes Isidore/Lili (2002), Gustav/Ike (2008) and Sandy (2012), respectively.}

\begin{tabular}{ p{2cm}  p{1.3cm}  p{1.2cm}  p{2cm}   p{2cm}   p{5cm} } 
\hline 
\multirow{2}{*}{\textbf{Hurricane}} & \multicolumn{2}{c}{\textbf{Date}}                                 & \multicolumn{2}{c}{\textbf{Transect coordinates}} &  \makecell{\multirow{2}{*} {\textbf{Transect location}}} \\ 

                  & \textbf{Before} & \textbf{After} & \centering\textbf{From} & \centering\textbf{To} & \\ 
\hline 
Isidore/Lili, 2002 & 15/09 & 07/10 & 21$^{\circ}$56$^\prime$15$''$N;  84$^{\circ}$56$^\prime$15$''$W   & 21$^{\circ}$56$^\prime$15$''$N;  86$^{\circ}$08$^\prime$45$''$W   & \multirow{3}{*}{\textbf{\includegraphics[scale=0.29]{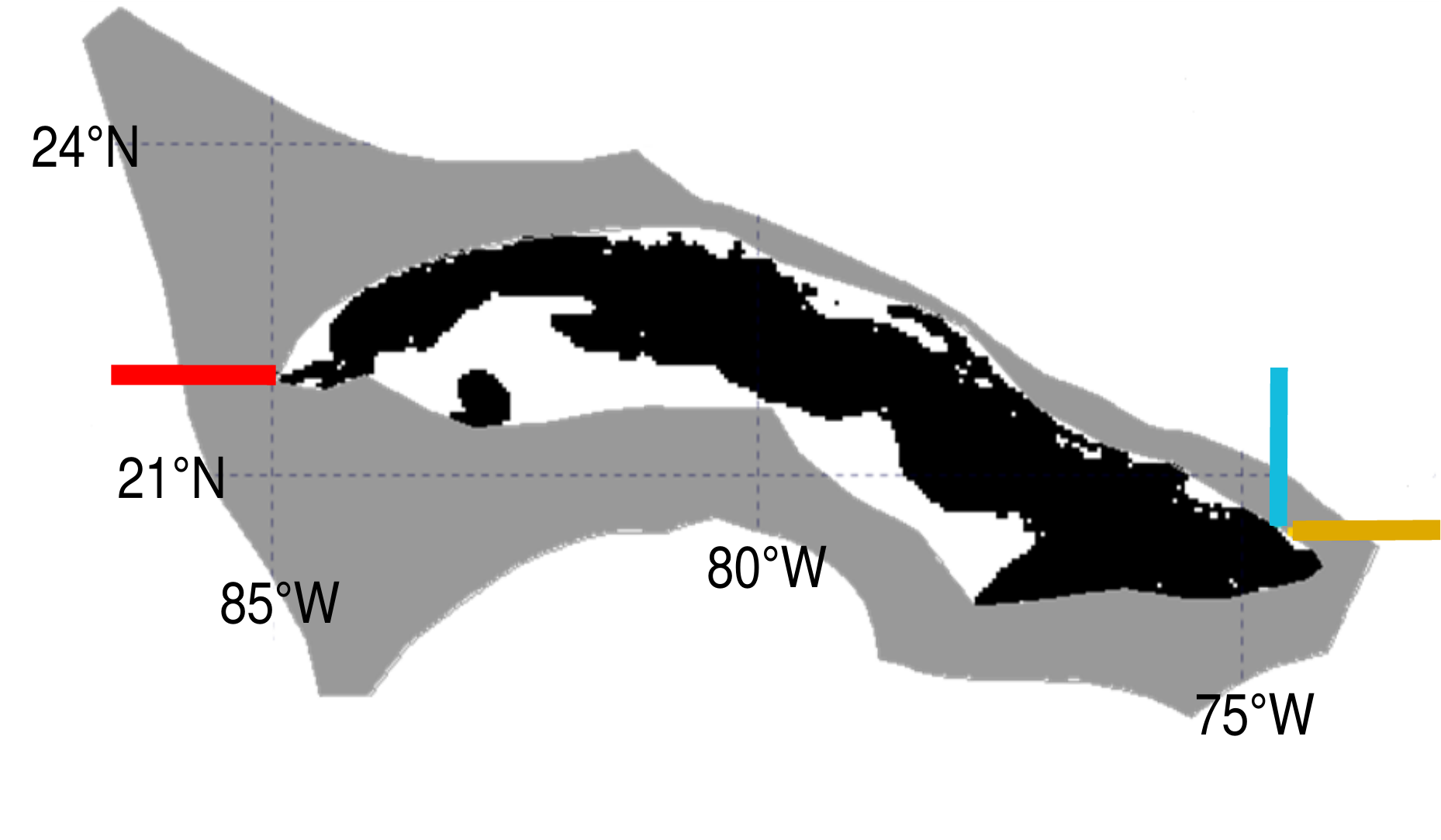}}} \\

 Gustav/Ike, 2008 & 22/08 & 12/09 &  20$^{\circ}$28$^\prime$45$''$N;  74$^{\circ}$28$^\prime$45$''$W   & 20$^{\circ}$28$^\prime$45$''$N;  73$^{\circ}$16$^\prime$15$''$W   &  \\

 Sandy, 2012 & 20/10 & 05/11 & 20$^{\circ}$33$^\prime$45$''$N;  74$^{\circ}$36$^\prime$15$''$W   & 21$^{\circ}$46$^\prime$15$''$N;  74$^{\circ}$36$^\prime$15$''$W   &    \\ 
 \hline 
 
 \end{tabular}  
\end{table}

Transects extended over 30 pixels since this number of pixels allowed to capture the spatial variability of the chl-a concentration for the selected bloom events. Although the used chl-a products have a spatial resolution of 0.0417$^{\circ}$ $\times$ 0.0417 $^{\circ}$ (i.e., $\sim$4.6~km at equator), we considered each pixel as a sector of 4~km $\times$ 4~km. We discarded the first two pixels (i.e., $\sim$0--4~km and $\sim$4--8 km) since they still contained biased information associated with the optically complex waters adjacent to the baseline of the EEZ for the analysed cases. Given the square shape of pixels, they are georeferenced at their center point~\mbox{\citep{gove2013}}. Thus, we considered the distance from the baseline to the center point of each pixel, i.e., 10~km (118~km) from the baseline to the center point of the first (last) pixel analysed. We confirmed that chl-a increases after the passage of hurricanes were related to the presence of phytoplankton given the consistent spatial variability of the mean chl-a and a$_{ph}$ as was analysed in~\cite{acker2009} (Figure~8). 

In agreement with \mbox{\cite{gove2016}}, we found that the relationship between the mean chl-a concentration (and also the mean a$ _{ph} $ in our case) and the distance from the baseline of the EEZ was best described by a power law ($\mathrm{\textit{p}}<0.05$, $\mathrm{R}^2=0.92$) (Figure~8). We used the slope of the linear regression equation fitted to log-log transformed data (log chl-a vs log distance to the baseline) as a proxy for the IME magnitude, as such enabling the comparison of chl-a concentration gradients among the study locations~\mbox{\citep{gove2016}}. The magnitude of the IME changed following the passage of hurricanes (as evidenced by differences in the slopes, Figure~9), in the sense that it was 89$\%$ higher after the passage of hurricanes. The outer boundary of the EEZ intersects the transects for Isidore/Lili (2002), Gustav/Ike (2008) and Sandy (2012) at approximately 98, 70 and 50~km, respectively, and the post-storm chl-a concentrations at these intersection points were 5, 44, and 74$\%$ higher, respectively, than those at the end of the transects. Given the variable width of the EEZ, those percentages of chl-a increase vary spatially, i.e., the narrower the EEZ the higher the value of chl-a concentration at the intersection point of the transect and the outer boundary of the EEZ as compare with the one at the end of the transect. Still, considering the analysed cases, it may be stated that the post-storm IME of Cuba leads to an organic fertilization of the adjacent waters of the Caribbean Sea. The IME magnitudes are lower than the ones reported for several islands and attols in the Pacific Ocean by~\cite{gove2016} (i.e., a slope ranging from 0 to --0.4). These authors based their analysis on the depth-integrated chl-a concentration (i.e., 5--300~m) using ship-based surveys. In contrast, remotely sensed chl-a concentrations provide an estimate of phytoplankton biomass in the upper ocean~\citep{gordon1975}, though it increases deeper in the water column, reaching a subsurface maximum that can be far greater than that observed in surface waters~\citep{furuya1990}. 

\subsection{Regression analyses}
Taking into account that the largest differences between the scales considered in this study were observed during the first two post-storm weeks, and given that the oceanographic conditions in the third and fourth post-storm weeks were influenced by other meteorological phenomena (see further in Discussion), the regression analyses were performed using the mean value of the first 15~days after the passage of hurricanes. A similar methodology, considering temporal and spatial means, has been previously used to assess the ocean response to TCs~\citep{babin2004, foltz2015, Parkeretal2017}. The determination coefficients of both the simple and multiple linear regression analyses are shown in Table~3. We found that the SST decrease at both scales after the passage of hurricanes is mainly determined by wind speed, and to a lesser extent by DAP. Together, these explanatory variables account for 56 and 54$\%$ ($\mathrm{\textit{p}}<0.05$) of the SST variability depending on the scales (Table~3). The SST cooling induced by rains might have been observed during the days before the hurricane passage given the high DAP anomalies at this time (Figure~3B). Yet, though it has been reported that SST starts to decrease at day --3 \mbox{\citep{vincent2012a, menkes2016}}, we found that this occurred only from the day~0 on (Figure~3C). Thus, the DAP effects did not seem to have been sufficient to cool the waters of the EEZ. This cooling only occurred under the combined effects of the highest wind speed and DAP at the moment of the passage (Figures~3A~and~3B).

For what concerns the chl-a concentration, we obtained large differences between the spatial scales. The chl-a relationships with wind speed and SST at scale~1~and~2 account for 32 and 58$\%$ of the chl-a variability, respectively, being only statistically significant ($\mathrm{\textit{p}}<0.05$) at scale~2 (Table~3). Whereas, wind speed and SST were likely the main drivers of the post-storm chl-a response at scale~2, other environmental factors appear to influence the chl-a responses at scale~1. It has been argued that an elevated chl-a concentration after the passage of TCs is a consequence of the entrainment of subsurface chl-a-rich waters and/or the new production due to the input of nutrients from below the nitracline~\citep{walker2005, gierach2008, menkes2016}. Although the largest chl-a responses ($\geq$ 0.1 mg m$^{-3}$) occurred for DNO$_{3}$ shallower than 60~m (delineated by the dashed contour in Figure~10A), just as found by~\cite{menkes2016}, we obtained a very weak correlation between the chl-a concentration and the DNO$_{3}$ (Table~3).

\begin{table}[H]
\centering
\caption{Determination coefficients of the simple and multiple linear regression analyses. Bold numbers indicate that the linear regression was significant ($\mathrm{\textit{p}}<0.05$). Transit speed of hurricanes and DNO$_{3}$ were not included in the multiple linear regression analyses given their low values of the determination coefficient for the simple linear regression analyses.}
\begin{tabular}{p{6.5cm}p{2.3cm}p{2.3cm}} 
\hline 
\multicolumn{3}{c}{\textbf{Simple linear regression}} \\ 
\hline 
\textbf{Explanatory variables of SST} & \textbf{Scale 1} & \textbf{Scale 2} \\ 

Wind speed & \textbf{0.515} & \textbf{0.489} \\ 

DAP & 0.207 & 0.157 \\ 

Transit speed & 2x10$^{-4}$ & 0.023 \\ 

\textbf{Explanatory variables of chl-a} & \textbf{Scale 1} & \textbf{Scale 2} \\ 

Wind speed & \textbf{0.242} & \textbf{0.375} \\ 

DAP & 0.139 & 1.63x10$^{-5}$ \\ 

Transit speed & 0.072 & 0.007 \\ 

SST & \textbf{0.267} & \textbf{0.563} \\ 

DNO$_{3}$ & 0.002 & 0.018 \\ 

Total coastal rainfall & 0.142 & 1.52x10$^{-5}$ \\ 
\hline 
\multicolumn{3}{c}{\textbf{Multiple linear regression}}\\ 
\hline 
\textbf{Explanatory variables of SST} & \textbf{Scale 1} & \textbf{Scale 2} \\ 

Wind speed-DAP & \textbf{0.560} & \textbf{0.540} \\ 
 
\textbf{Explanatory variables of chl-a} & \textbf{Scale 1} & \textbf{Scale 2} \\ 

SST-Wind speed & 0.320 & \textbf{0.583} \\ 

SST-Wind speed-DAP & \textbf{0.810} & \textbf{0.677} \\ 

SST-Wind speed-Total coastal rainfall & \textbf{0.670} & \textbf{0.621} \\ 
\hline 
\end{tabular} 
\end{table}

Although the relationship between the chl-a concentration and the DAP at scale~1 is not statistically significant (Table~3), they seem to be negatively correlated as shown in Figure~10B. This result disagrees with~\cite{lin2016} who reported that heavy rainfall enhances blooming after the passage of typhoons in the western North Pacific. The post-storm chl-a enhancement mechanism proposed by~\cite{lin2016} could negatively impact the chl-a response after the passage of TC in areas with a limited entrainment of subsurface nutrients. In such areas, the limited amount of upwelled nutrients (and/or chl-a) would be trapped by the strong surface halocline (formed by the rainfall-induced surface salinity decrease). These nutrients will be rapidly assimilated by the phytoplankton, thereby depleting the surface layer, as the halocline hinders mixing and hence nutrient replacement. In contrast, low DAP values could prevent the formation of a strong halocline, which in turn can allow a continued entrainment of subsurface nutrients and/or chl-a, resulting in a moderate increase of the surface chl-a concentration. This could have been the case for the waters of the EEZ of Cuba, as such explaining the negative relationship found between chl-a concentration and DAP along the trajectory.

Although the wind speed, SST and DAP explain 81$\%$ ($\mathrm{\textit{p}}<0.05$) of the chl-a concentration variability at scale~1 (Table~3), the latter only appears to regulate the effects induced by the wind speed and SST. To identify other environmental factors directly inducing the chl-a response at this scale, we collected the total amount of rainfall in stations near the coastline of Cuba (taken from the TC reports of the NHC) at those days that the hurricanes affected Cuba, and we considered them as a proxy of the coastal rainfalls in the first post-storm week. As the TCs make landfall, the coastal land is inundated. This heavy rainfall can induce large river discharges and subsequent high concentrations of nutrients in coastal waters, leading to a high phytoplankton production. Then, local oceanic currents can horizontally transport those waters to the open ocean. We only used the data of total coastal rainfall, because it has been reported that it induces a stronger chl-a response for larger periods after the passage of the hurricanes, than for example wind speed in coastal regions~\citep{angles2015, Parkeretal2017}.

In Figure~10C we can observe the relationship between chl-a concentration and the total amount of coastal rainfall at scale~1. In general, the determination coefficients obtained for the regressions between chl-a and its explanatory variables individually at scale~1 were low, which could indicate that several processes influence the observed chl-a response. The combined effects of wind speed, SST and total coastal rainfall account for 67$\%$ ($\mathrm{\textit{p}}<0.05$) of the chl-a variability at scale~1, while the addition of the latter variable at scale~2 does not considerably improve the percentage of variability explained by wind speed and SST (Table~3). Given that~13~of the studied hurricanes made landfall in Cuba, the coastal phytoplankton production could have been stimulated through rainfall-induced biological response. \cite{farfan2014} reported that the relative increase in chl-a concentration following TC landfalls in Yucatan and Baja California Peninsulas appeared to be strongly linked to the intensity of precipitation in coastal regions. The remaining percentage of the chl-a concentration variability that was not captured by the multiple linear regression is probably due to other environmental factors (not considered in this study), such as the chl-a transport from phytoplankton bloom regions in the Gulf of Mexico. 

On the other hand, transit speed appears to be irrelevant in determining the post-storm SST and chl-a responses at both considered scales (Table~3). Slower-moving TCs have relatively more time to input momentum into the upper ocean, which eventually produces more sea surface cooling~\citep{zedler2009, mei2012} and an increased surface chl-a concentration~\citep{gierach2008}. However, slower-moving hurricanes studied here (e.g., Paula (2010), Mathew (2016), Katrina (2005) and Rita (2005)) (Table~1) had low transit time values within the EEZ (i.e., 12~h, Table~1) and they generated, on average, weak SST (--0.5$^{\circ}$C) and chl-a (0.025 mg m$^{-3}$) responses during the first two post-storm weeks. We found that the transit distance and time of the analysed hurricanes inside the EEZ of Cuba show a strong relationship ($\mathrm{\textit{p}}<0.05$, $\mathrm{R}^2=0.89$), so, a large transit distance implies a long transit time (Table~1) and this can explain the obtained weak correlations of the post-storm SST and chl-a with transit speed. In contrast, the transit distance and time appear to be more important in determining the chl-a response at scale~1 ($\mathrm{\textit{p}}<0.05$, transit distance $\mathrm{R}^2=0.135$, transit time $\mathrm{R}^2=0.252$) as was reported by~\cite{babin2004}. The post-storm SST responses do not show a clear dependence with the above-mentioned metrics. 

\section{Discussion}
\subsection{Wind speed dynamics}
The maximum mean wind speed anomaly in our study occurs at day +1 (Figure~3A), as 11~hurricanes attained their greatest wind speed that day. Moreover, 6 hurricanes reached a higher SSHS at day +1, while 3 hurricanes did this beyond day +1, but still inside the EEZ. This increase in SSHS and hence wind speed can be explained by the relationship between SST and hurricane intensification~\citep{fraza2015, arora2016}. Although it has been demonstrated that the SST decreased at day +1 (i.e., 0.25 and 0.21$^{\circ}$C at scale~1~and~2, respectively), as compared to the SST at day~0, the SST at day~+1 was still high with minimum and maximum values of 27.6 and 29.6$^{\circ}$C, respectively, across all hurricanes and both scales. Such SSTs allow for the intensification of hurricanes, as was confirmed by~\cite{fraza2015} for North Atlantic hurricanes. They reported that temperatures of 27.6$^{\circ}$C or higher will lead to a high hurricane intensification~\cite [see Figure 7 in] [] {fraza2015}. In general, the Gulf of Mexico and the western Caribbean Sea are regions of maximum TC rapid intensification (i.e., maximum sustained surface wind speed increase of at least 15.4~m~s$^{-1}$ over a 24~h period) in the North Atlantic \mbox{\citep{wang2017}}. Specifically, the highest climatological TC rapid intensification in the North Atlantic agrees with the highest climatological SST and TC heat potential (integrated vertical temperature from the sea surface to the depth of the 26$^{\circ}$C) in the waters surrounding Cuba~\cite [see Figure 2 in] [] {wang2017}. Nevertheless, the wind speed of some hurricanes did not increase, while others even weakened in the EEZ, but it is known that other environmental factors can limit the TC intensity~\citep{whitney1997}.

\subsection{DAP dynamics}
The high positive mean DAP anomalies in the pre-storm week (Figure~3B) might have been caused by the asymmetric TC rainfall distribution. For the Atlantic Basin the mean maximum precipitation intensity of all TCs in the period 1998--2000 was located in the front-right quadrant of TCs up to a distance of 300~km from their center~\citep{lonfat2004}. Moreover, the maximum intensity increases with the TC translation speed, with fast moving TCs ($>$5~m~s$^{-1}$) having a rain rate as high as 50$\%$ of the ambient mean. This could explain the high DAP anomalies in the pre-storm week, since parts of the studied areas could have been significantly affected by the rainfall from the front-quadrants of the hurricanes, given that 50$\%$ of the studied hurricanes were fast moving according to the criterion of~\cite{lonfat2004}. Besides, at days~0 and +1, high DAP positive anomalies were recorded, as compared to the ones of the pre-storm week (Figure~3B), possibly because many of the studied hurricanes intensified within or when they left the EEZ. Over oceans, it has been shown that the TC rainfall amount is correlated with the storm maximum wind speed, though a lag in the maximum rainfall amount and maximum wind speed has been reported~\citep{jiang2008}. The maximum mean rain rate near the center of the hurricanes appears to precede approximately 2 days the maximum surface winds~\cite [reviewed by] [] {jiang2008}. Thus, the hurricanes that intensified in the EEZ should have caused heavy rainfalls some days before their center reached the EEZ (i.e., at day~--2), while the hurricanes that intensified after leaving the EEZ, should have shown a high DAP within the EEZ. For example,~\cite{jiang2008} reported that the rapid intensification of Lili (2002) on early 3 October in the Gulf of Mexico was related to a strong convective burst on 1 October, i.e., the day it reached the Cuban EEZ. 

The pronounced temporal variability of the mean post-storm DAP anomalies (Figure~3B) has a multifactorial origin. The variability during the first and the second post-storm weeks accounts for the effects of the post-storm DAP of the studied hurricanes. The DAP peaks during those weeks result from the pre-storm rainfalls of Lili (2002) and Ike (2008), and the post-storm rainfalls of Paloma (2008), which weakened to a tropical depression and became nearly stationary over Cuba,            (\textit{http://www.nhc.noaa.gov/data/tcr/AL172008$\textunderscore$Paloma.pdf}) leaving the EEZ only 6 days after its entrance. In contrast, the temporal variability in the third and the fourth post-storm weeks was influenced by the effects of other meteorological phenomena. 

The Caribbean rainfall season spans from May to November, and is of bimodal nature~\citep{taylor2002}. Its early rainfall season begins in May and lasts until June, with a brief dry period in July, while the late rainfall season spans from August to November~\citep{taylor2002}. As we can see, almost the entire rainy season coincides with the hurricane season, hence, TCs are not the only climatological phenomenon contributing to the rainfall in the region. For example, during the fourth post-storm week of Paloma (2--9 December of 2008) high DAP values were recorded. According to the Monthly Synoptic Summary of the National Institute of Meteorology of Cuba, December 2008 was affected by six cold fronts (with the highest activity occurring during the first ten days of the month) and almost two thirds of these cold fronts were preceded by rainfall (\textit{www.met.inf.cu}), which increased the DAP in the EEZ. Besides, a high DAP was recorded during the third post-storm week of Ivan (2004) and Ike (2008). In these cases, the distant effect of hurricane Jeanne (2004) (not analysed here) induced rainfalls in the northwestern of the EEZ and hence impacted the DAP values in the third post-storm week of Ivan (2004), while effects of easterly tropical waves, and low-pressure systems played a role in determining the DAP values in the third post-storm week of Ike (2008) as was reported in the Monthly Synoptic Summary of the National Institute of Meteorology of Cuba (\textit{www.met.inf.cu}).

\subsection{SST dynamics}
The limited sea surface cooling reported in this study agrees with the results of \cite{menkes2016}. In the absence of deep warm ocean waters (e.g., like the ones of the warm core rings), the resulting post-storm SST decrease typically ranges from 3 to 5$^{\circ}$C \citep{meyers2014}. However,~\cite{uhlhorn2012} reported a SST decrease less than 1$^{\circ}$C after Lili's passage over the Loop Current. Besides, in the Gulf of Mexico~\cite{meyers2016} found that a wind-forced entrainment of colder thermocline water into the ocean mixed layer (OML) caused a SST decrease of 5$^{\circ}$C, while SST decreased only 1--2$^{\circ}$C in the Loop Current after the passage of Gustav/Ike (2008). Although we have no direct evidence to explain the limited cooling, it might be a consequence of the thermal structure of the waters of the EEZ of Cuba. 

The deep warm Loop Current influences the southwestern and northwestern parts of the EEZ. This current transports the very warm waters of the Caribbean Sea, which in turn are characteristic of the entire southern region of the EEZ. In the waters transported by the Loop Current, the stratification is usually weak and water of 26$^{\circ}$C can be found at depths greater than 100 m~\citep{meyers2014}, while in subtropical waters, such as those of the Caribbean Sea, the 20$^{\circ}$C isotherm can be found between 180 and 220 m of depth~\citep{shay2000}. So, the water column of the EEZ of Cuba is warm down to great depths. 

Besides, the SST decrease after the passage of TCs is related with the depth of the OML. If the OML is shallow, i.e., there is cold water near the surface, then the wind-induced mixing generates a more significant surface cooling~\citep{meyers2016}. In general, a very deep mixed layer has been linked to a particularly weak SST cooling under the eye of the TC~\citep{shay2000}. \cite{muller2015} reported that the mean summer OML (for the period 1992--2012) in a quadrant enclosing the northwestern waters of the EEZ of Cuba was slightly thicker ($\sim$30~m), as compared to other areas in the Gulf of Mexico ($\sim$20~m). Figure~11 indicates that the OML before the passage of Ivan (2004), Katrina (2005) and Gustav (2008) was deeper in the western areas of the EEZ than in some regions of the Gulf of Mexico. So, this could have limited a strong TC-induced surface cooling in the study area, and hence have implied a weak chl-a response. 

The progressive post-storm cooling (on the basis of the mean SST anomalies) observed in Figure~3C disagrees with the results of~\cite{menkes2016}. Although they reported that cooling persists for at least 30~days after the TC passage, they found a considerable decrease of the cooling occurring from day~+2 onward. Our results can be explained, however, by the fact that the post-storm conditions of some hurricanes were the pre-storm conditions of the following one in the season. In such cases, a net decrease of the SST occurred, even though the SST in the EEZ was sufficiently high to fuel the following hurricane of the season. The mean pre-storm SST along the trajectories of Katrina, Rita and Wilma (2005) was 30.17, 29.64 and 28.8$^{\circ}$C, respectively (Figure~12A). So, there was a decrease of about 0.5$^{\circ}$C between the pre-storm weeks of Katrina and Rita and a decrease of 0.84$^{\circ}$C between the pre-storm weeks of Rita and Wilma. Similar observations were made for the hurricanes of 2002 and 2008 and at scale~2 for all these hurricanes (Figure~12B). 

The pre-storm weeks of the above-mentioned hurricanes were in the months of August and September. In August, the waters of the Gulf of Mexico reach their maximum SST (warmer than 29.5$^{\circ}$C), and by September the warm pool expands southward into the Caribbean Sea and eastward into the western tropical North Atlantic~\citep{wang2007}. Given that the EEZ of Cuba encloses parts of the Gulf of Mexico and the Caribbean Sea, it is expected that during these months the waters of the EEZ remain warm. Consequently, the observed SST decrease can be attributed to the hurricanes. Nevertheless, the more negative anomalies in the fourth post-storm week were caused by the seasonal cooling phase in November and December, coinciding with the post-storms weeks of many of the analysed hurricanes. This agrees with the findings of~\cite{price2008} for Jeanne (2004) in the North Atlantic. This also explains the low SSTs in the fourth post-storm week of Michelle (2001), Wilma (2005), Paloma (2008) and Tomas (2010) as compared with the other hurricanes (Figures~12A~and~12B).

\subsection{Chl-a dynamics}
The positive mean pre-storm chl-a anomalies (Figure~3D) were caused in part by the horizontal transport of chl-a from other regions of the Gulf of Mexico. For example, during the pre-storm week of Charley (2004) the TC Bonnie passed as a tropical depression across the western part of the EEZ. Bonnie continued to the Gulf of Mexico, where it intensified to a tropical storm and made landfall near St.\ Vincent Island (\textit{http://www.nhc.noaa.gov/data/tcr/AL022004$\textunderscore$Bonnie.pdf}). This allowed the transport of a high-chl-a plume from the Mississippi River Delta through the Loop Current system (Figure~13A). Besides, during the entire pre-storm week of Paula (2010) a chl-a increase to the north of the Yucatan Peninsula was observed (Figure~13B). This led to a high pre-storm chl-a concentration as compared to the post-storm concentration of the above-mentioned hurricanes at both scales analysed here (Figures~12C~and~12D).

The high chl-a concentration anomalies in the first and the second post-storm weeks (Figure~3D) agree with previous studies reporting blooms lasting about 2--3 weeks after the TC passage~\citep{babin2004, hanshaw2008, Parkeretal2017}. The first chl-a peak immediately after the passage agrees with the findings of~\cite{lin2016} and~\cite{menkes2016}. The latter observed weaker chl-a responses (i.e., 60$\%$ of the responses range between --0.02 and 0.02 mg m$^{-3}$) than the ones in our study at scale~1 (i.e., 60$\%$ of the responses range between --0.01 and 0.04 mg m$^{-3}$) (Figure~4C). This discrepancy could be due to the different satellite chl-a products used. Moreover, \cite{menkes2016} assessed the chl-a response in disks of 200~km radius centered at the TC positions, while we used disks of 120~km. Since larger blooms occur near the TC trajectory \citep{babin2004, walker2005, gierach2008, lin2016}, a smaller scale of analysis implies a stronger chl-a response.

On the other hand, the high chl-a concentration anomalies at the end of the second post-storm week (Figure~3D) reflect the combined effects of Isidore and Lili (2002), where the second week after the passage of Isidore showed a very high chl-a concentration (i.e., 0.17 mg m$^{-3}$) (Figures~12C~and~12D). These hurricanes followed similar trajectories in the EEZ and were only 9~days apart. A similar case was the one of Gustav and Ike (2008), though the chl-a responses of these were more limited than the one caused by Isidore (2002). Although the trajectories of Ike and Gustav (2008) crossed in the northwest of the  EEZ, they followed different routes (Figure~2) and were 12~days apart. The peak chl-a anomalies in the fourth post-storm week are determined by multiple factors. For instance, changes of the SST due to the seasonal cycle, variability of precipitation because of cold fonts and other meteorological phenomena, as well as the chl-a transport from other regions of the adjacent LMEs. 

\subsection{Drivers of SST and chl-a responses}

The similarity of the determination coefficients for the simple and multiple linear regression of the relationship between the SST and its explanatory variables at both scales (Table~2) indicates that the same explanatory variables drove, in general, the SST response. Besides, acknowledging the similar distribution of the SST anomalies at both scales (Figures~4A~and~4B) allows us to conclude that the SST responses to hurricanes can be captured at these scales. TC-induced cooling affects wider regions and extends more into the open ocean~\citep{menkes2016}. Specifically, surface cooling usually can be observed up to 150--350~km from the hurricane center~\citep{cione2000}. In agreement,~\cite{vincent2012b} found that mean SST anomalies of approximately 1$^{\circ}$C can be observed typically over 5~times the radius of the maximum wind, i.e., 200~km beyond the TC position.

Reports indicate that the sources of chl-a contributing to the post-storm response can change between TCs~\citep{gierach2008, shropshire2016}. Distinguishing between the chl-a sources is crucial if we seek to understand the impact of storms on surface ocean biology \citep{shropshire2016}. The waters at scale~1 have a climatological mean DNO$_{3}$ of 57~m, which are deeper than the ones estimated by~\cite{shropshire2016} for the Gulf of Mexico and the Sargasso Sea (i.e., 35 and 5~m, respectively). So, entrainment of chl-a from the subsurface DCM could have dominated the vertical processes determining the post-storm chl-a response, corroborating the results of~\cite{shropshire2016}. The DCM occurs at the boundary between the euphotic layer and the top of the nitracline, so both have a similar spatial distribution in TC-prone regions~\citep{menkes2016}. Still, at some locations along the trajectories, nutrients could have reached the ocean surface and stimulated new phytoplankton production. Indeed, as the DNO$_{3}$ along the trajectory varies between a depth of 10 to 208~m, locations with the shallowest DNO$_{3}$ could have experienced both vertical processes, i.e., entrainment of subsurface nutrients and chl-a, allowing a higher and prolonged chl-a response after the TC passage~\citep{babin2004}.  

Both processes mentioned above result from the vertical transport of chl-a and/or nutrients from the deep ocean to the surface and may be considered as the most important mechanisms driving the biological response in open oceans. However, near coastal zones the responses could have been influenced by additional mechanisms. Given the obtained correlation of the transit distance and time with chl-a (not SST) at scale~1, it can be thought that post-storm chl-a concentration is affected by other processes rather than the vertical ones. Taking into account the results shown in Figure~6 and the relationship between the chl-a concentration and total coastal rainfall at scale~1 (Figure~10C), we argue that transport of rich coastal waters to the deep waters of the EEZ is important in explaining the post-storm chl-a variability along the hurricane trajectory. An overview of the mechanisms leading to an increase of the chl-a concentration in ocean waters near to the coastal areas is shown in Figure~14.
 
Filaments transporting coastally upwelled waters to the deep oligotrophic ocean in all eastern boundary currents of the subtropical gyre have been observed~\citep{strub1991}. Those filaments commonly have high nutrient concentrations near the coast, decreasing as the distance from the shore increases, which indicates that nutrients are utilised as they are transported~\citep{pelegri2005}. This spatial nutrient gradient matches the changes in phytoplankton community structure as one moves away from the coast. Indeed, a diatom-dominated assemblage commonly occurs near the coast, while a picoplankton community is abundant in the offshore portion of the filaments~\citep{joint2001}. Given this nutrient consumption, direct export of nutrients by such filaments to the open ocean may be limited, at least compared with the transport of organic matter (e.g., phytoplankton cells)~\citep{pelegri2005}. Extrapolating these findings to the filaments formed after the TC passage, extending from the coast to the oceanic waters of the EEZ of Cuba, entrainment of horizontally transported chl-a may be considered as the dominant horizontal mechanism determining the chl-a increases in the open ocean after the TC passage.

\section{Conclusions}
This study is a first attempt to assess the response of the waters of the EEZ of Cuba to the passage of hurricanes. During the period 1998--2016, 18~hurricanes crossed the EEZ, of which 12 were classified as major hurricanes at some point of their trajectories. Overall, those hurricanes induced a weak cooling of the ocean surface waters since 75 and 85$\%$ of the SST anomalies at scale~1~and~2, respectively, were smaller than --1$^{\circ}$C. The cooling was mainly caused by the wind, inducing mixing and upwelling of subsurface cool waters. The maximum chl-a responses were recorded in the first and second post-storm weeks, with 60$\%$ ranging between --0.01 and 0.04~mg~m$^{-3}$ at scale~1, and between --0.07 and 0.02~mg~m$^{-3}$ at scale 2. The wind speed and SST largely governed the chl-a response at the latter scale (58$\%$ of the chl-a variability), while 67$\%$ of chl-a response at scale~1 is explained by the combined effect of wind speed, SST and total coastal rainfall. We considered that the transport of chl-a from the DCM and/or the rich coastal waters are the dominant mechanisms determining the chl-a response in the EEZ after the passage of hurricanes. The remaining variability of the chl-a could be due to the transport of chl-a from phytoplankton bloom regions of the Gulf of Mexico. We also reported and quantified the magnitude of the IME in the EEZ after the passage of the hurricanes. This phenomenon may have caused a fertilization of the waters of the Caribbean Sea. At both scales, we observed post-storm SST and chl-a responses, but the latter seemed to depend more on the scale. A large spatial extent may lead to an underestimation of the biological effects induced by hurricanes, mask the large variability of the induced responses, and hinder the identification of the environmental factors governing the local chl-a responses. This study provides a benchmark which can be used to compare against which changes in oceanographic conditions can be evaluated after the passage of hurricanes. Besides, our results can stimulate and guide future studies to understand how this variability can affect the marine fisheries in the Cuban EEZ.

\section*{Acknowledgments}
We thank Dr.\ Iv\'an Ernesto P\'erez Santos for kindly assisting us on some satellite images analysis and the Upper Ocean Dynamics Laboratory at the University of Miami-Rosenstiel School of Marine and Atmospheric Sciences for kindly permitting us to use their ocean mixed layer images. This work was supported by the Special Research Fund (BOF) of Ghent University, Belgium, Grant Code 01W03715.

\bibliography{Avila-AlonsoetalHurricanes}  
\bibliographystyle{apa}

\section*{Figures}

\begin{figure} [H]
\centering                                                             
\includegraphics[scale=1]{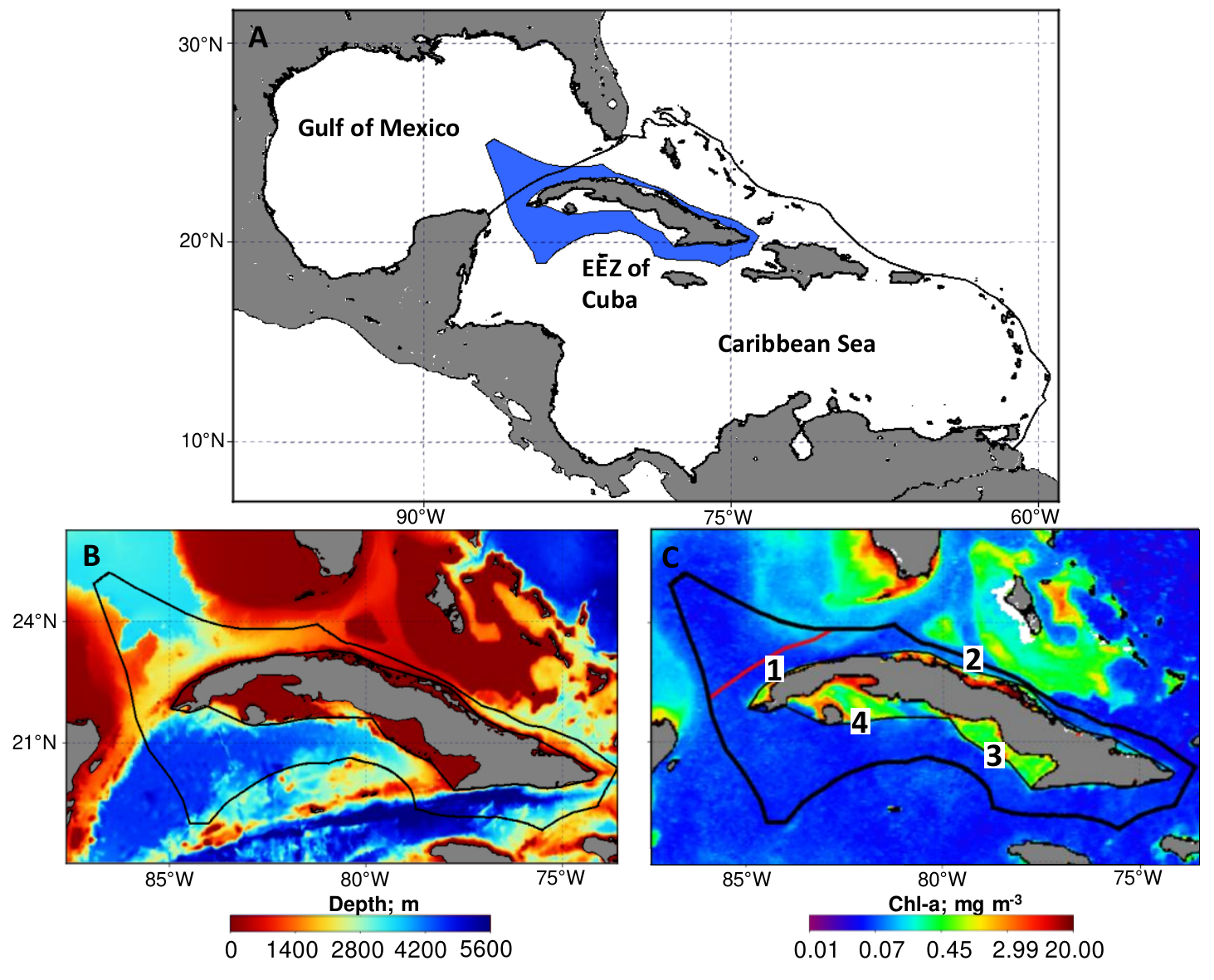}
\caption{(A) Exclusive Economic Zone (EEZ) of Cuba (blue area) and Large Marine Ecosystems (LMEs) Gulf of Mexico and Caribbean Sea. (B) Bathymetric map (ETOPO 1) of the study area. (C) 8 days composite chlorophyll-a (chl-a) image (20-27 August 2008, pre-storm week of the hurricane Gustav) indicating the four shallow macrolagoons close to the coastline of Cuba, i.e., (1) Los Colorados Archipelagos, (2) Sabana-Camag\"{u}ey Archipelagos, (3) Ana Maria and Guacanayabo Gulfs and (4) Batabano Gulf. The red line indicates the boundary between the LMEs.}
\end{figure}

\begin{figure} [H]
\centering 
\includegraphics[scale=0.7]{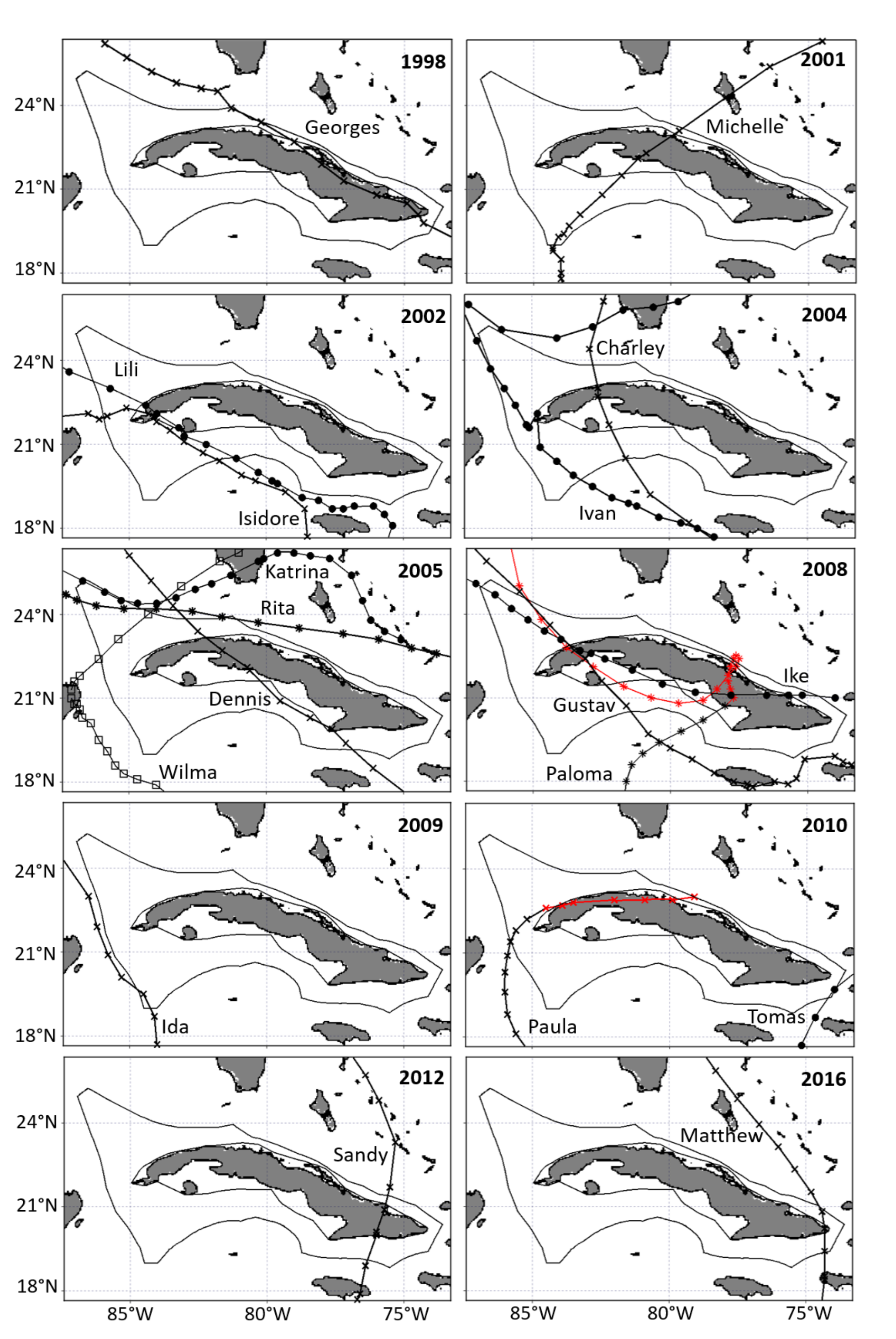}
\caption{Trajectories of the hurricanes affecting the Exclusive Economic Zone of Cuba in the period~1998--2016. The red trajectories of the hurricanes Paloma in 2008 and Paula in 2010 indicate that they weakened and became TCs of lower intensity.}
\end{figure}

\begin{figure}[H]
\centering 
\includegraphics[scale=0.76]{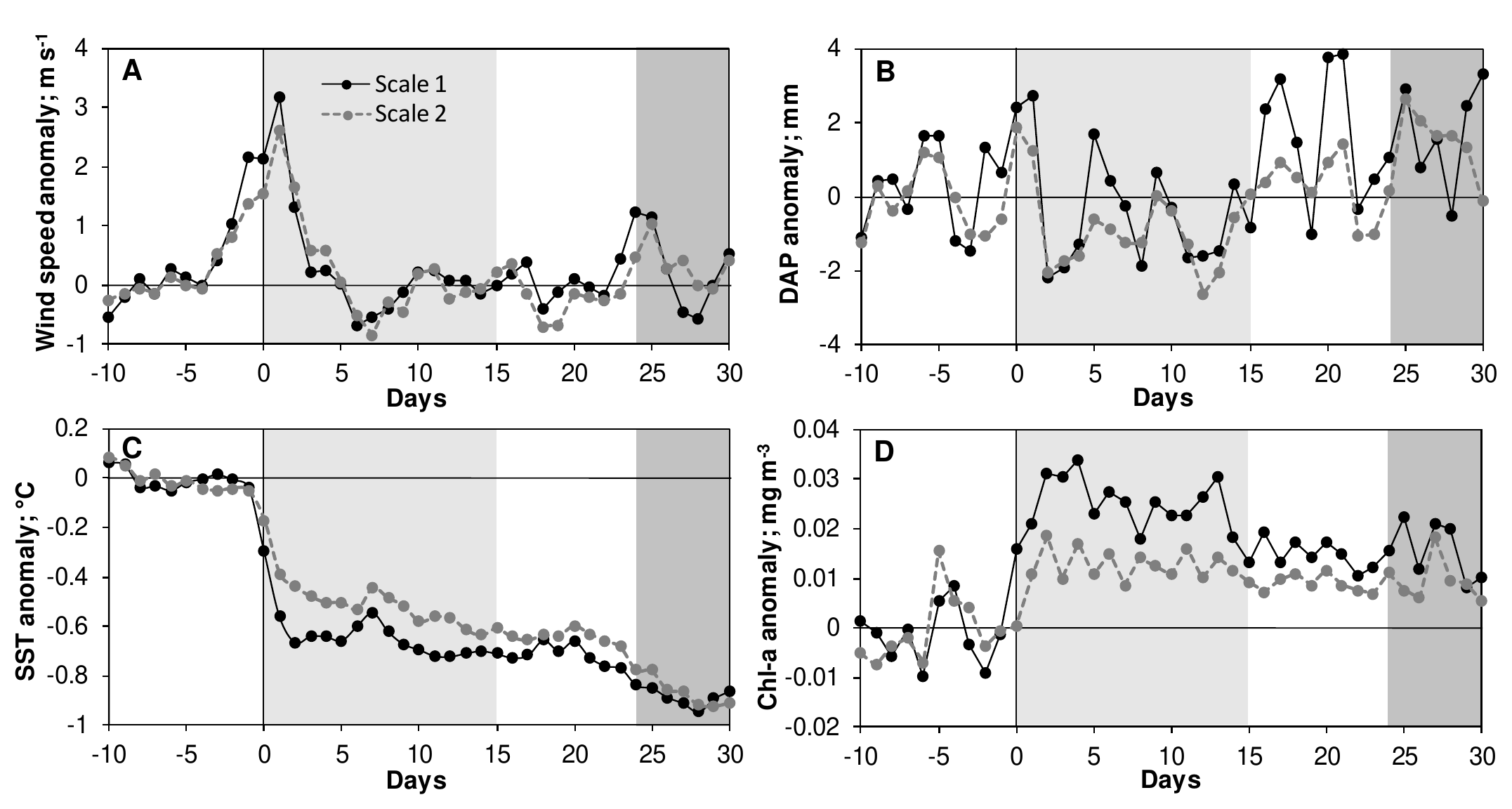}
\caption{Mean daily evolution of anomalies of (A) wind speed, (B) daily accumulated precipitation (DAP), (C) sea surface temperature (SST) and (D) chlorophyll-a (chl-a) concentration at scale~1 (along the trajectory of the hurricanes) and scale~2 (throughout the entire Exclusive Economic Zone of Cuba). Light grey areas indicate the first and second post-storm weeks, while the dark grey areas indicate the fourth post-storm week.}
\end{figure}

\begin{figure}[H]
\centering 
\includegraphics[scale=0.9]{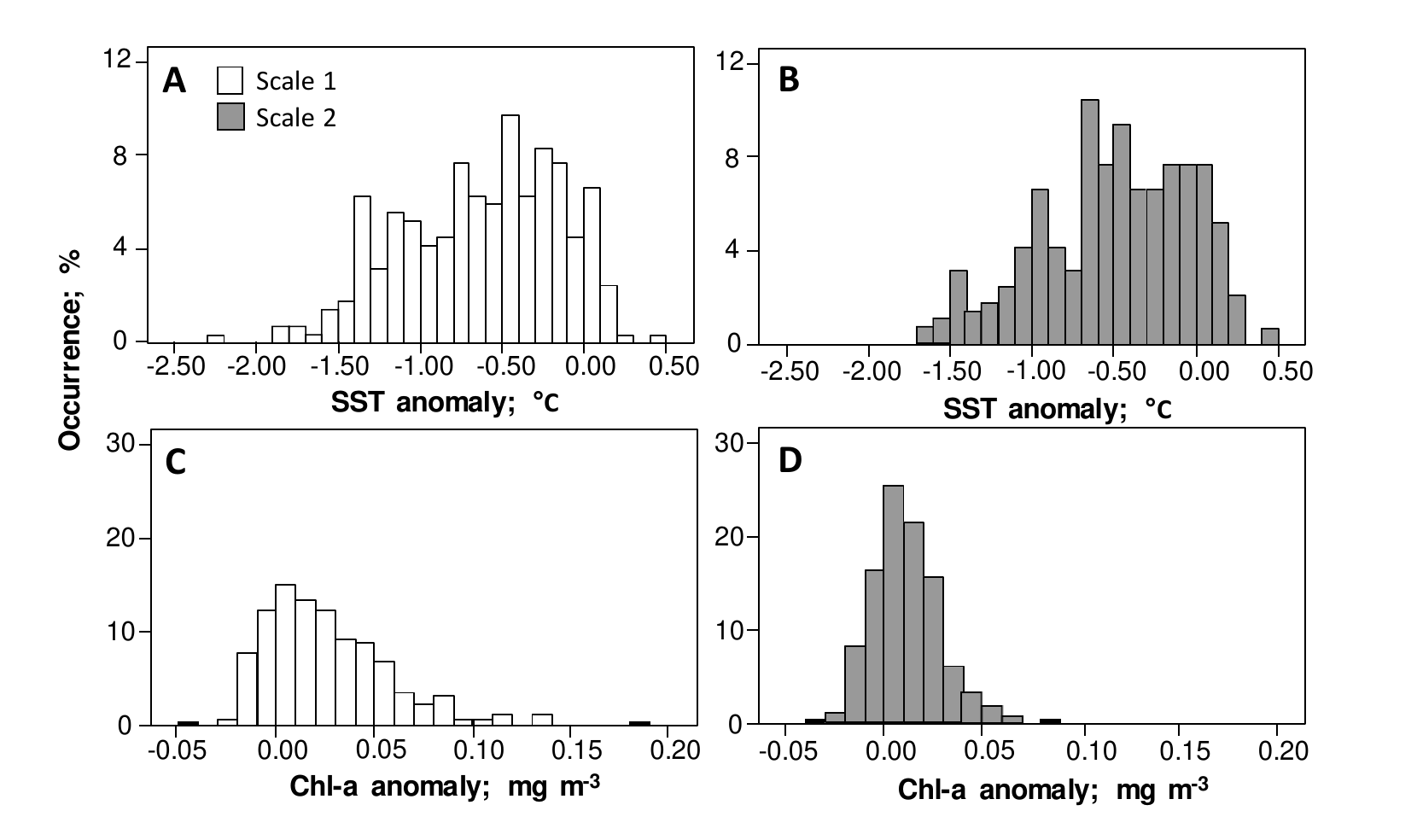}
\caption{Distributions of (A and B) sea surface temperature (SST) and (C and D) chlorophyll-a (chl-a) anomalies in the first and second post-storm weeks at (A and C) scale~1 (along the trajectory of the hurricanes) and (B and D) scale~2 (throughout the entire Exclusive Economic Zone of Cuba).}
\end{figure}

 \begin{figure}[H]
\centering 
\includegraphics[scale=0.9]{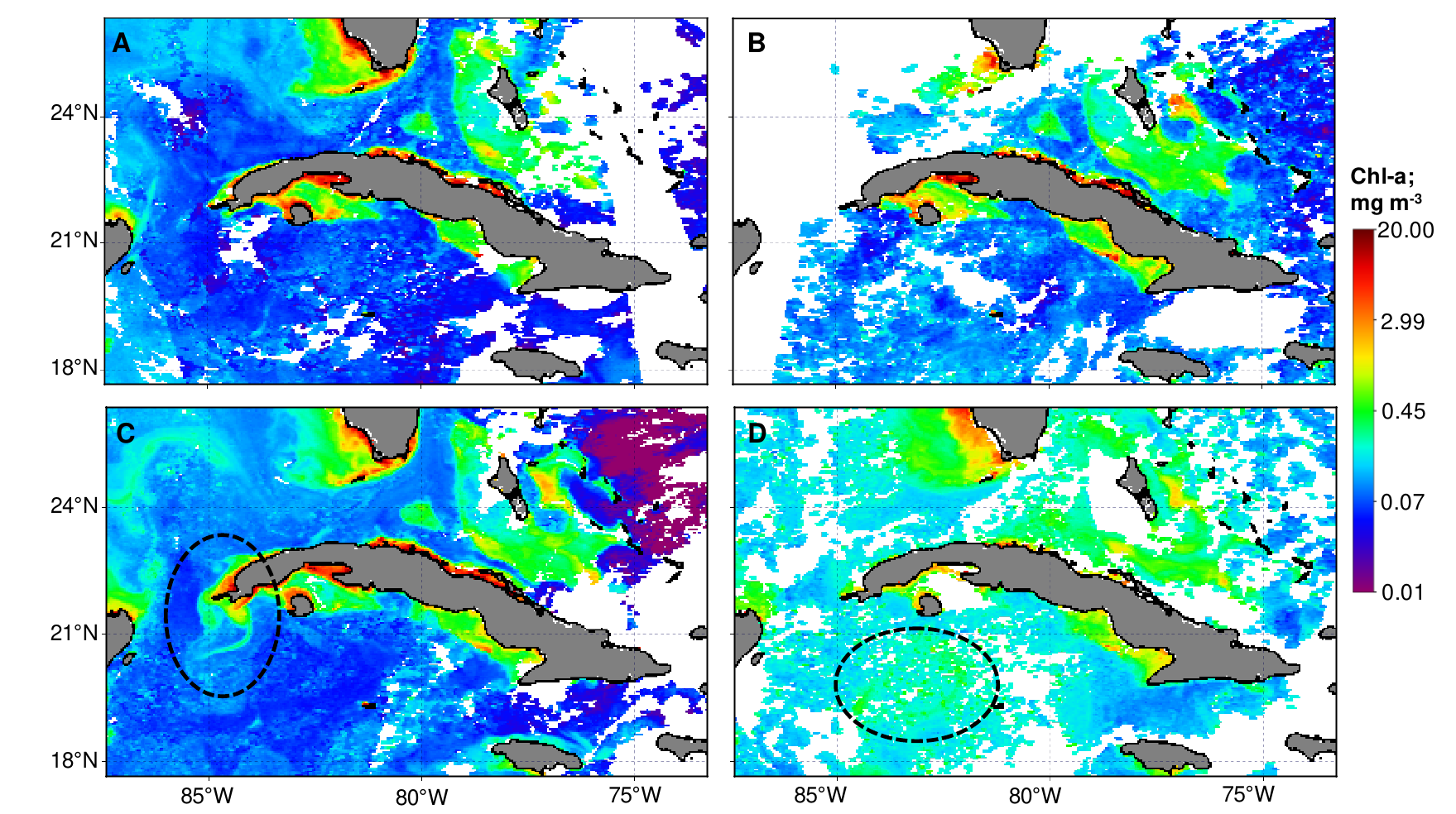}
\caption{Chlorophyll-a (chl-a) concentration before and after the passage of hurricanes: Representative pre-storm days of (A) Isidore/Lili (September 15, 2002) and (B) Michelle (October 26, 2004). Local chl-a increases (indicated by the dashed contours) after (C) Isidore/Lili (October 4, 2002) and (D) Michelle (November 7, 2001). In (D) also a  global chl-a increase throughout the entire EEZ is shown.}
\end{figure}

 \begin{figure}[H]
\centering 
\includegraphics[scale=0.9]{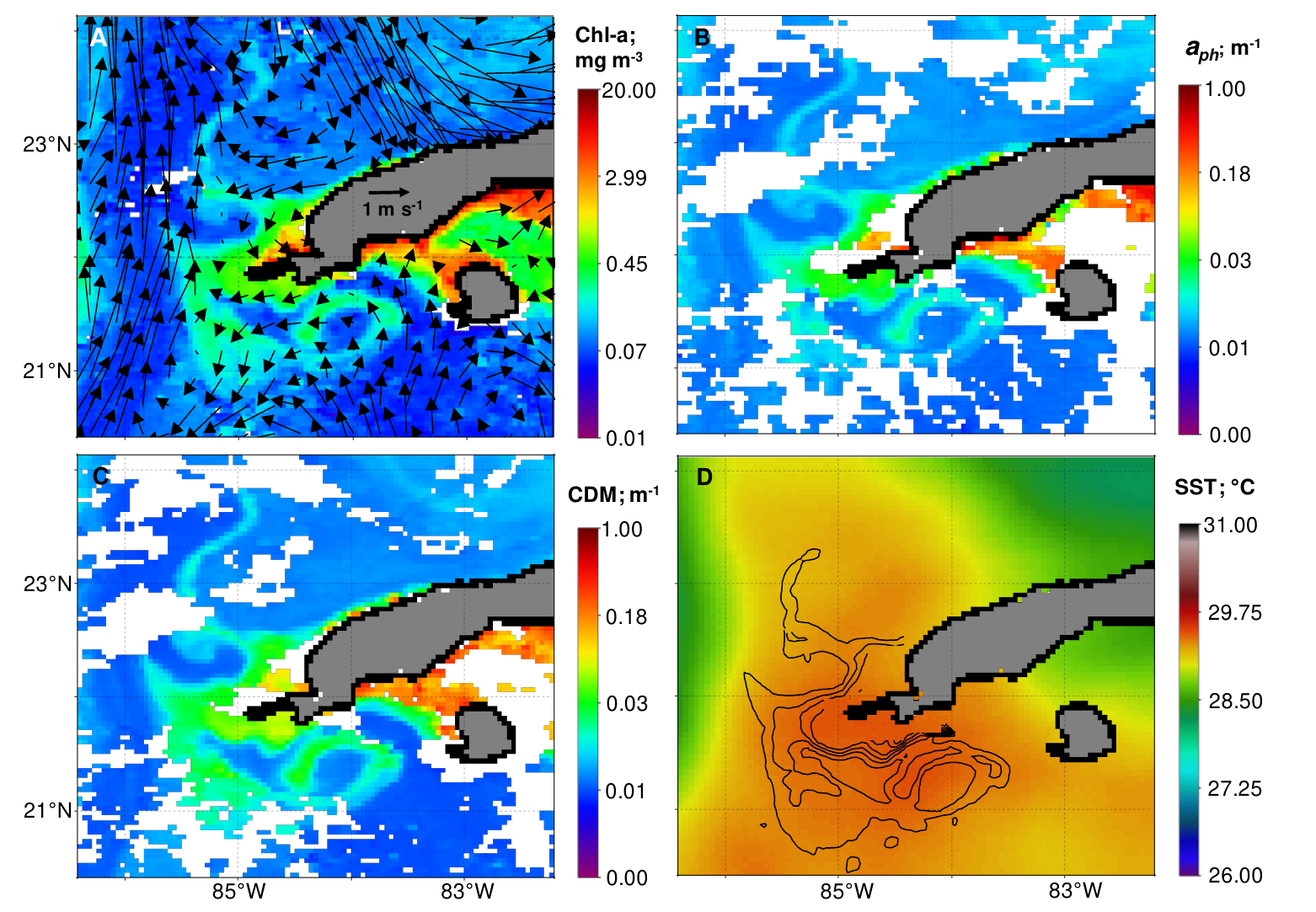}
\caption{(A) Local increase of chlorophyll-a (chl-a) concentration after the passage of Isidore/Lili (October 8, 2002) with the vector field of the geostrophic marine currents superimposed and the corresponding (B) phytoplankton absorption coefficient at 443~nm (a$ _{ph}$), (C) detrital organic materials absorption coefficient at 443~nm (CDM) and (D) sea surface temperature (SST) with a chl-a contour line of 0.13~mg~m$^{-3} $ superimposed.}
\end{figure}

\begin{figure}[H]
\centering
\includegraphics[scale=0.75]{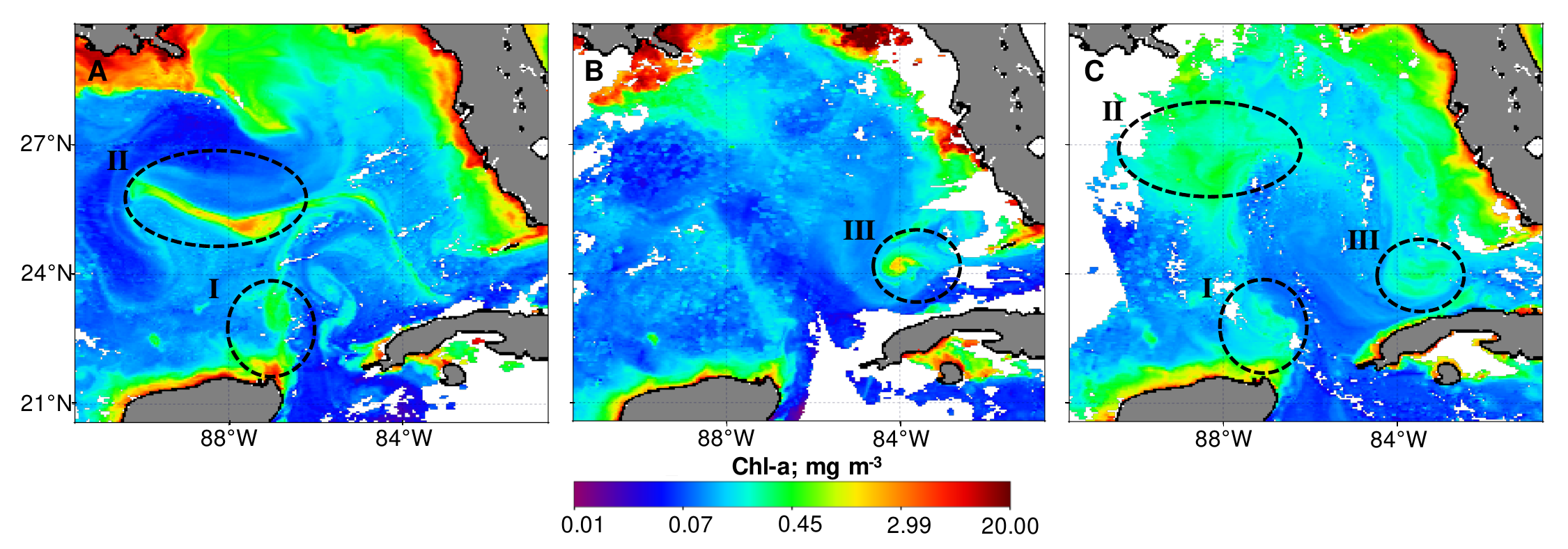}
\caption{Chlorophyll-a (chl-a) concentration in the Gulf of Mexico after the passage of (A) Ivan (September 18, 2004), (B) Katrina (August 31, 2005) and (C) Gustav/Ike (September 13, 2008). Dashed contours indicate regions I (eastern Campeche Bank), II (central Gulf of Mexico) and III (southwest of the Florida Peninsula) where chl-a increases were typically observed.}
\end{figure}

\begin{figure}[H]
\centering 
\includegraphics[scale=0.55]{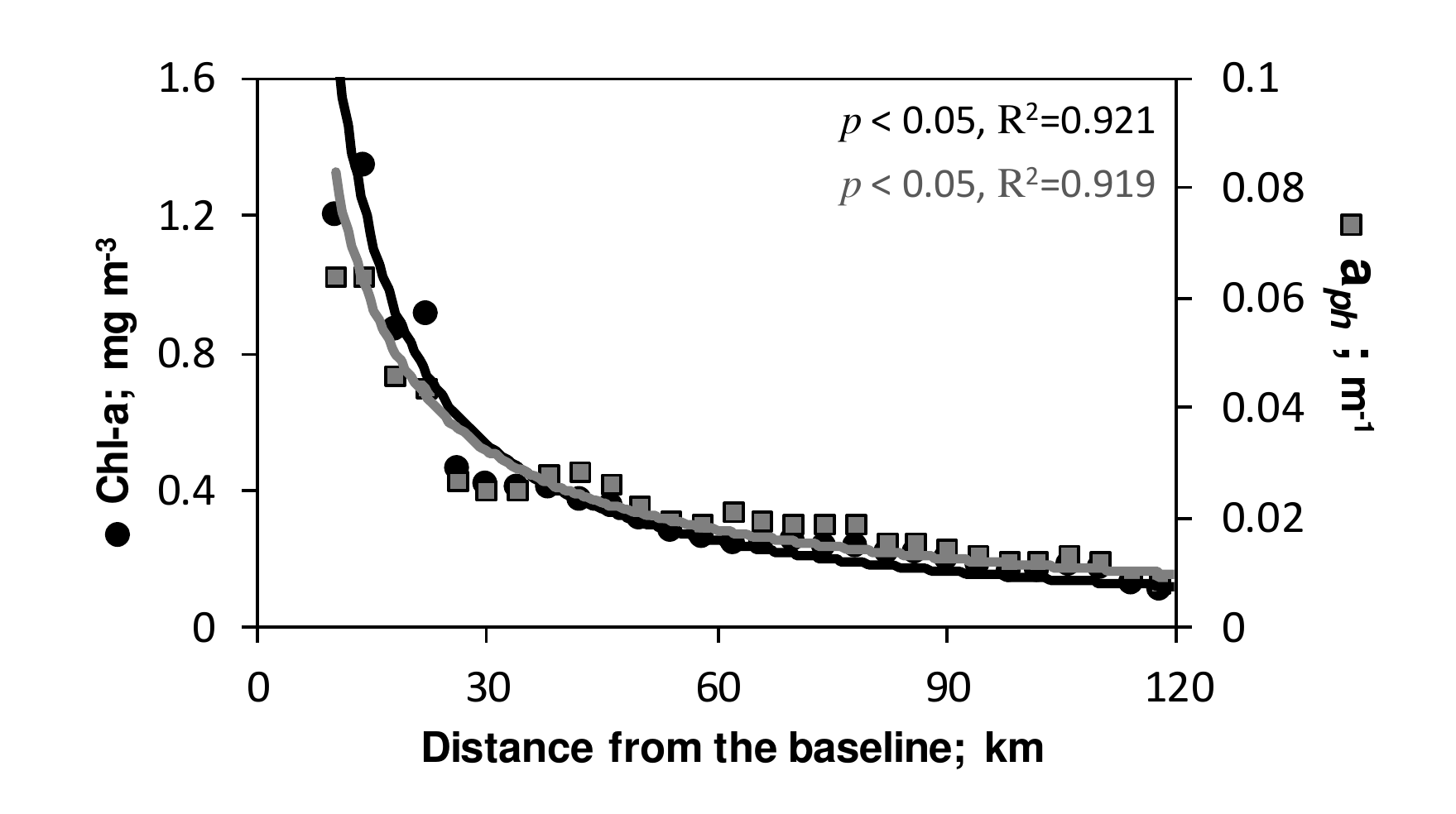}
\caption{Mean post-storm surface chlorophyll-a (chl-a) concentration and phytoplankton absorption coefficient at 443~nm (a$_{ph}$) versus the distance from the baseline of the Exclusive Economic Zone of Cuba.}
\end{figure}

\begin{figure}[H]
\centering
\includegraphics[scale=0.7]{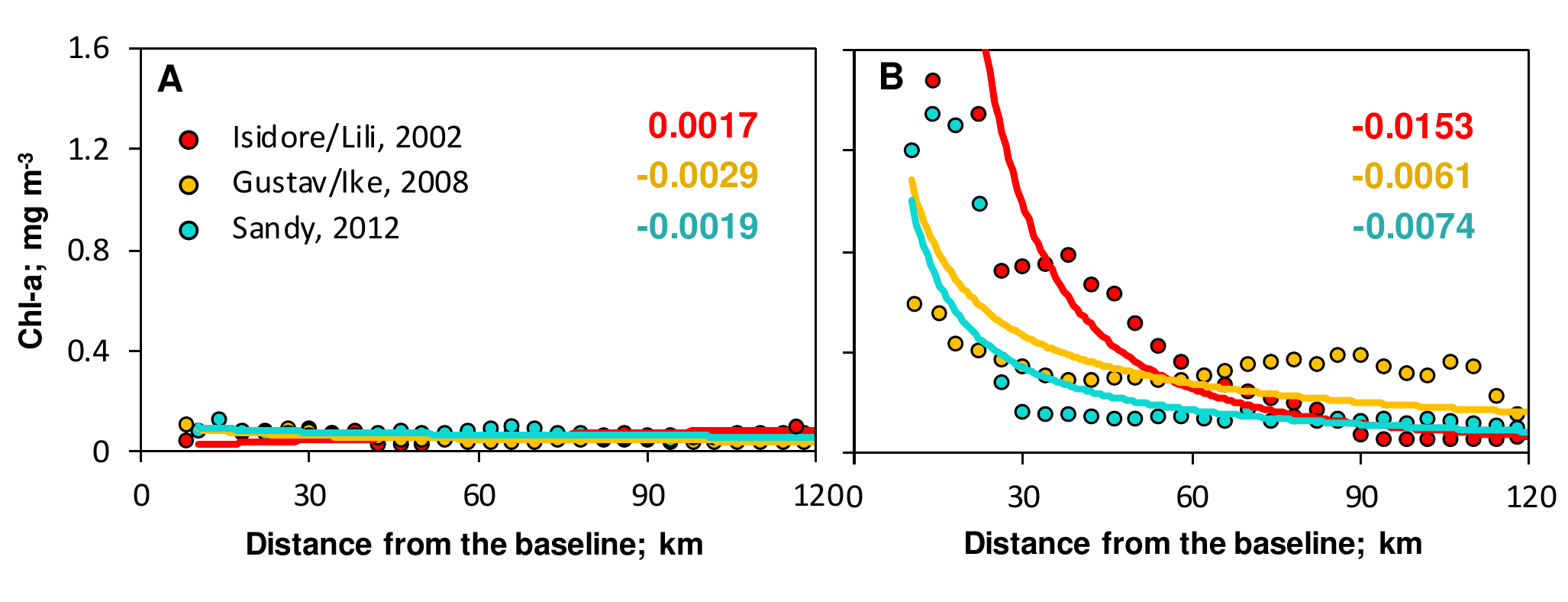}
\caption{Chlorophyll-a (chl-a) concentration versus the distance from the baseline of the Exclusive Economic Zone of Cuba (A) before and (B) after the passage of hurricanes in locations with a significant post-storm chl-a increase ($\mathrm{\textit{p}}<0.05$). Dots indicate remotely sensed chl-a data and solid lines represent the best fit function (power function). Nonlinear regressions ($\mathrm{\textit{p}}<0.05$) with R$^{2}$ values ranging from 0.25 to 0.91. The numbers in colour indicate the IME magnitude (i.e., slope of linear fit of log chl-a vs log distance from the shore).}
\end{figure}

\begin{figure}[H]
\centering
\renewcommand{\baselinestretch}{1.5}
\includegraphics[scale=0.75]{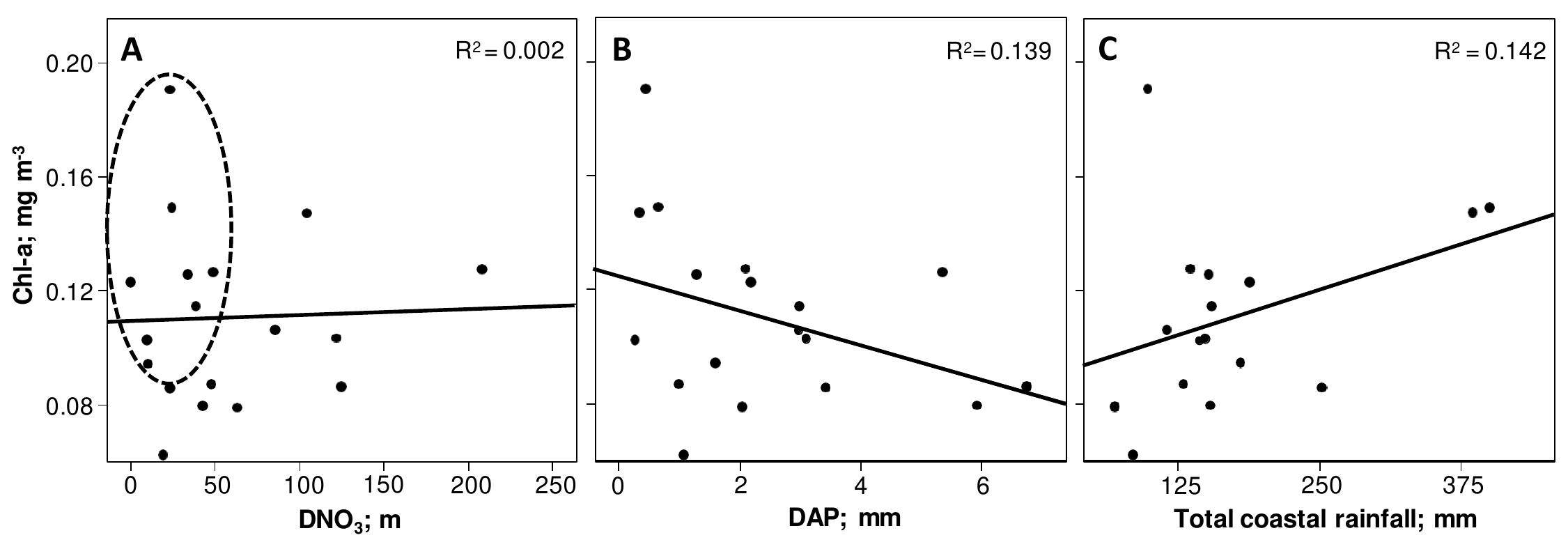}
\caption{(A) Chlorophyll-a (chl-a) versus nitracline depth (DNO$_{3}$), (B) daily accumulated precipitation (DAP) and (C) total coastal rainfall at scale 1 (along the trajectory of the hurricanes). The dashed contour in (A) indicates chl-a values equal and larger than 0.1~mg~m$^{-3}$ that occurred for DNO$_{3}$ shallower than 60~m.}
\label{Reg2}
\end{figure}

\begin{figure}[H]
\centering
\includegraphics[scale=0.7]{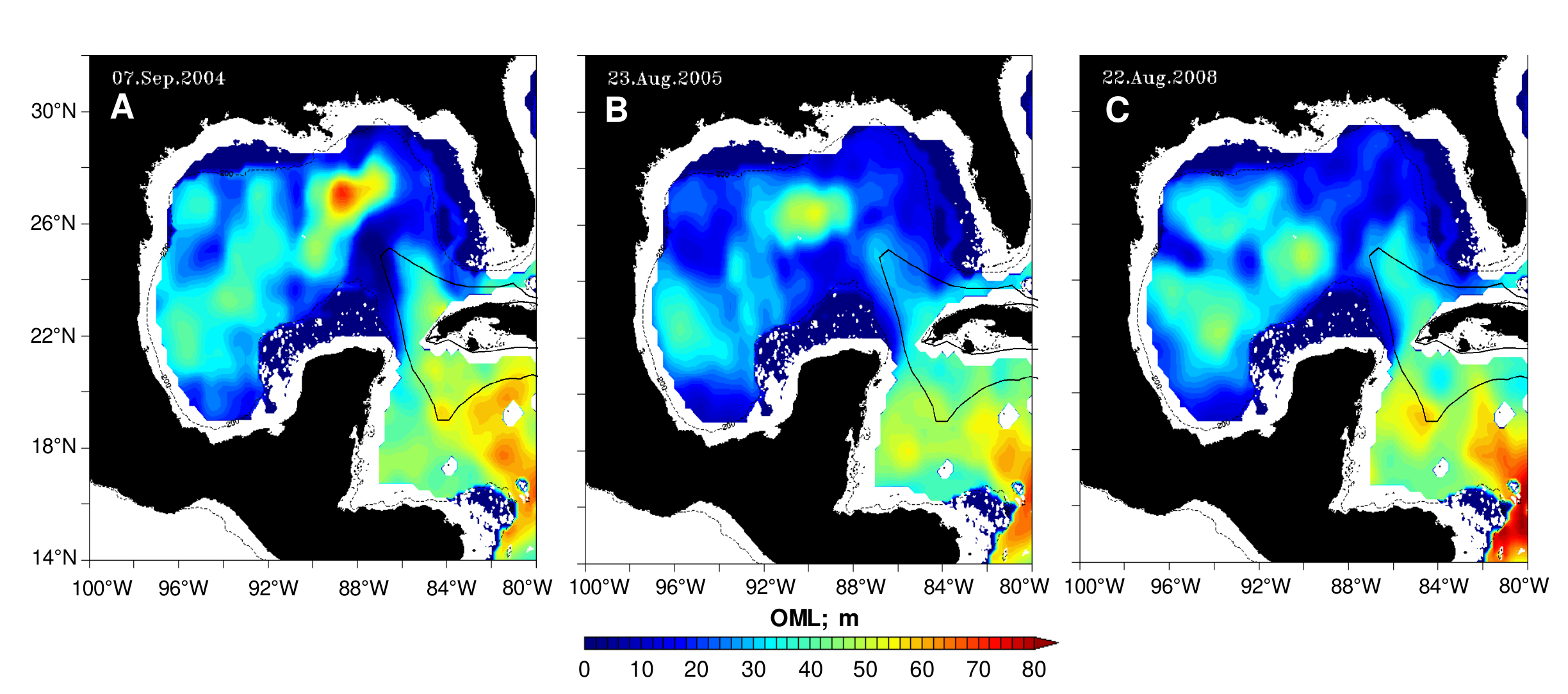}
\caption{Ocean mixed layer (OML) depth in the western Caribbean Sea and the Gulf of Mexico before the passage of (A) Ivan (2004), (B) Katrina (2005), and (C) Gustav (2008) across the Exclusive Economic Zone (EEZ) of Cuba. Superimposed is the boundary of the EEZ of Cuba. Images were provided by the Upper Ocean Dynamics Laboratory at the University of Miami Rosenstiel School of Marine and Atmospheric Sciences (\textit{www.rsmas.miami.edu/groups/upper-ocean-dynamics}). OML depth (0.25$^{\circ}$ $\times$ 0.25$^{\circ}$ of spatial resolution) was computed by using the Systematically Merged Atlantic Regional Temperature and Salinity Climatology to blend the Generalized Digital Environmental Model and the World Ocean Atlas 2001 climatologies \citep{meyers2014}.}
\end{figure}

\begin{figure}[H]
\centering
\includegraphics[scale=0.55]{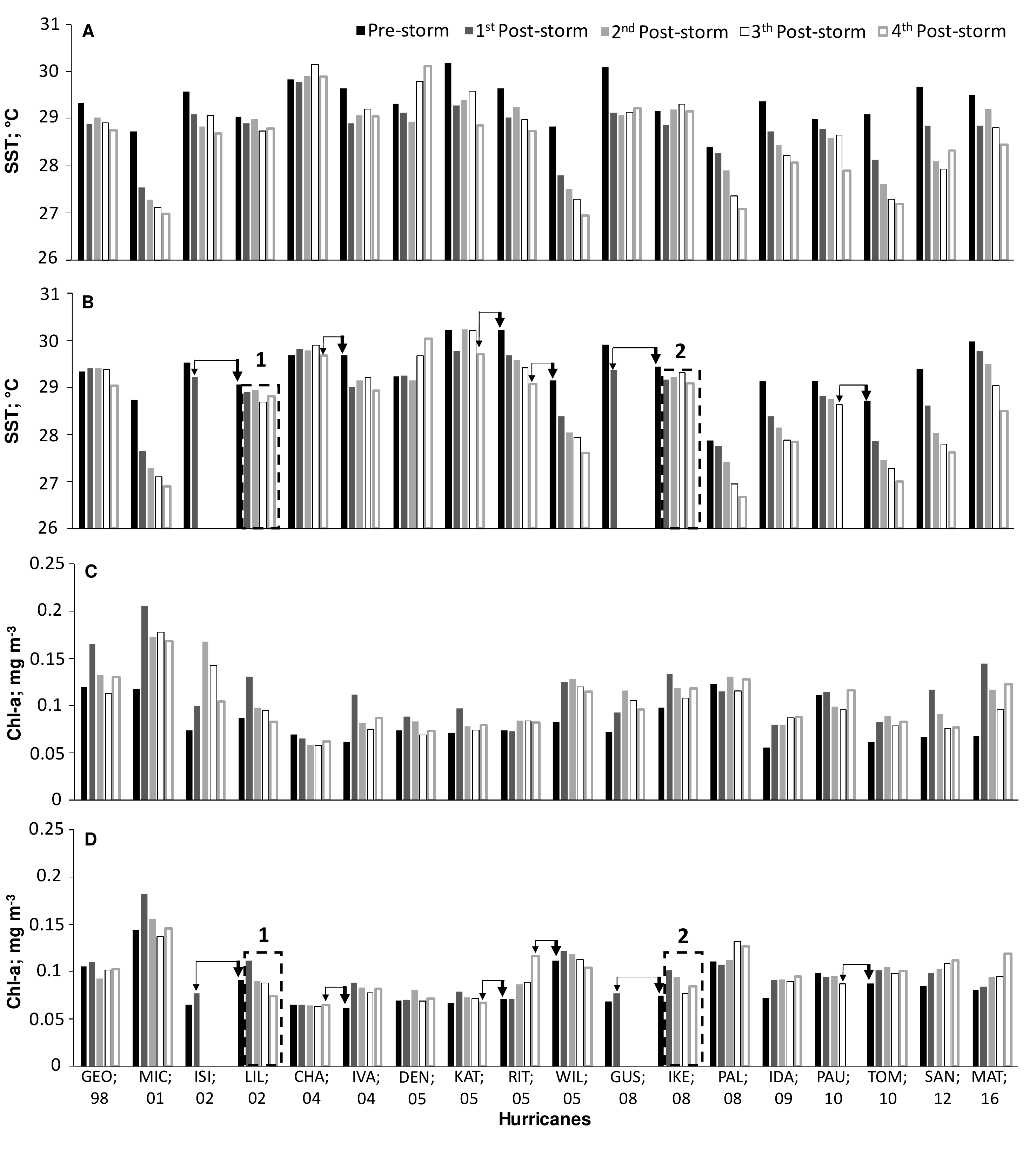}
\caption{Weekly mean of the pre- and post-storm sea surface temperature (SST) and chlorophyll-a concentration (chl-a) (A and C) at scale~1 (along the trajectory of the hurricanes) and (B and D) at scale~2 (throughout the entire Exclusive Economic Zone of Cuba (EEZ)). Thick arrows indicate that the pre-storm week is influenced by the post-storm week of the prior hurricane (indicated by a thin arrow). Dashed rectangles indicate combined hurricane effects: 1 (Isidore and Lili in 2002) and 2 (Gustav and Ike in 2008).}
\end{figure}

\begin{figure}[H]
\centering
\includegraphics[scale=0.8]{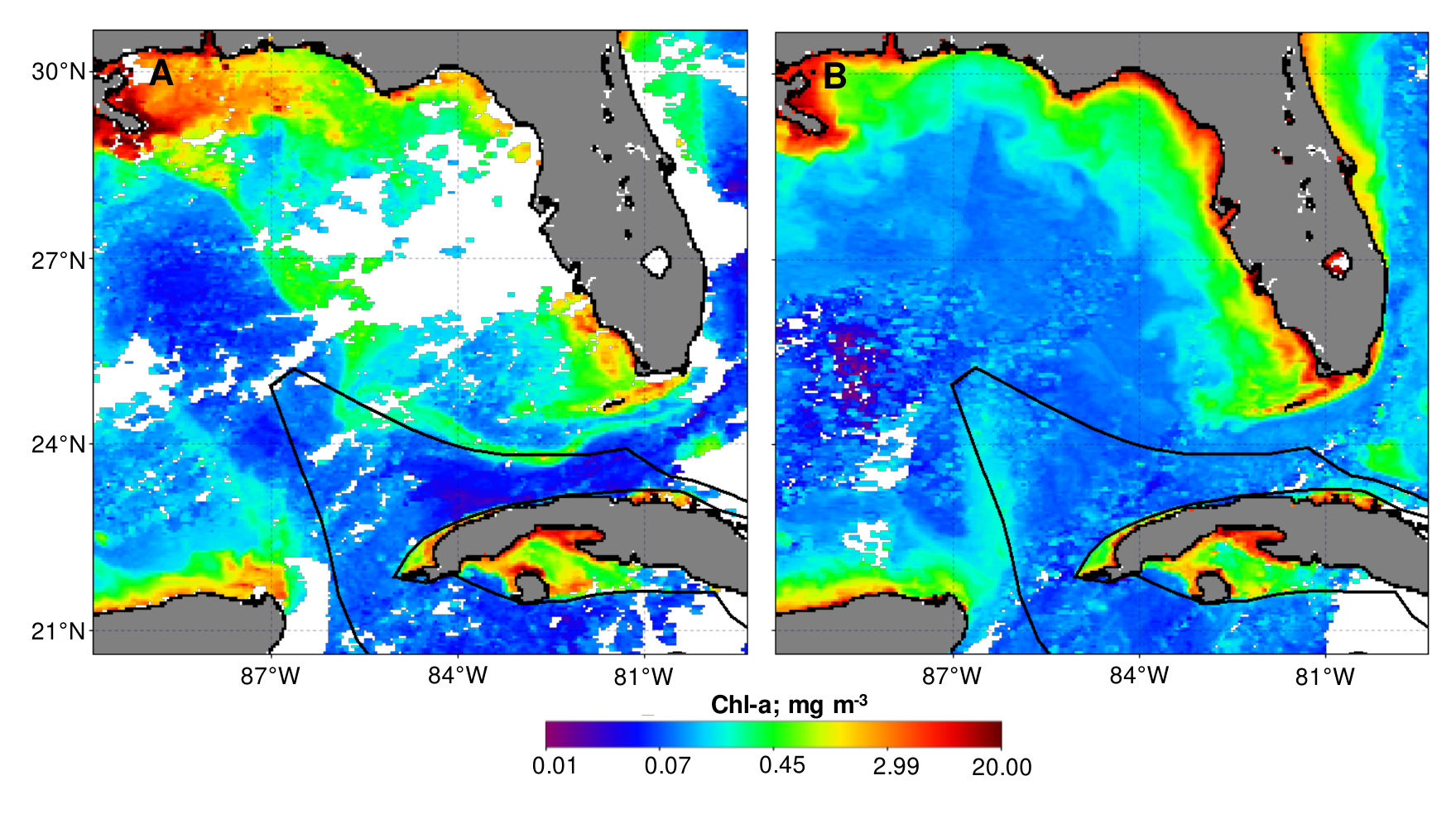}
\caption{Transport of chlorophyll-a (chl-a) from the Gulf of Mexico to the Exclusive Economic Zone of Cuba in the pre-storm week of (A) Charley (August 4, 2004) and (B) Paula (October 9, 2010).}
\end{figure}

\begin{figure}[H]
\centering
\includegraphics[scale=0.8]{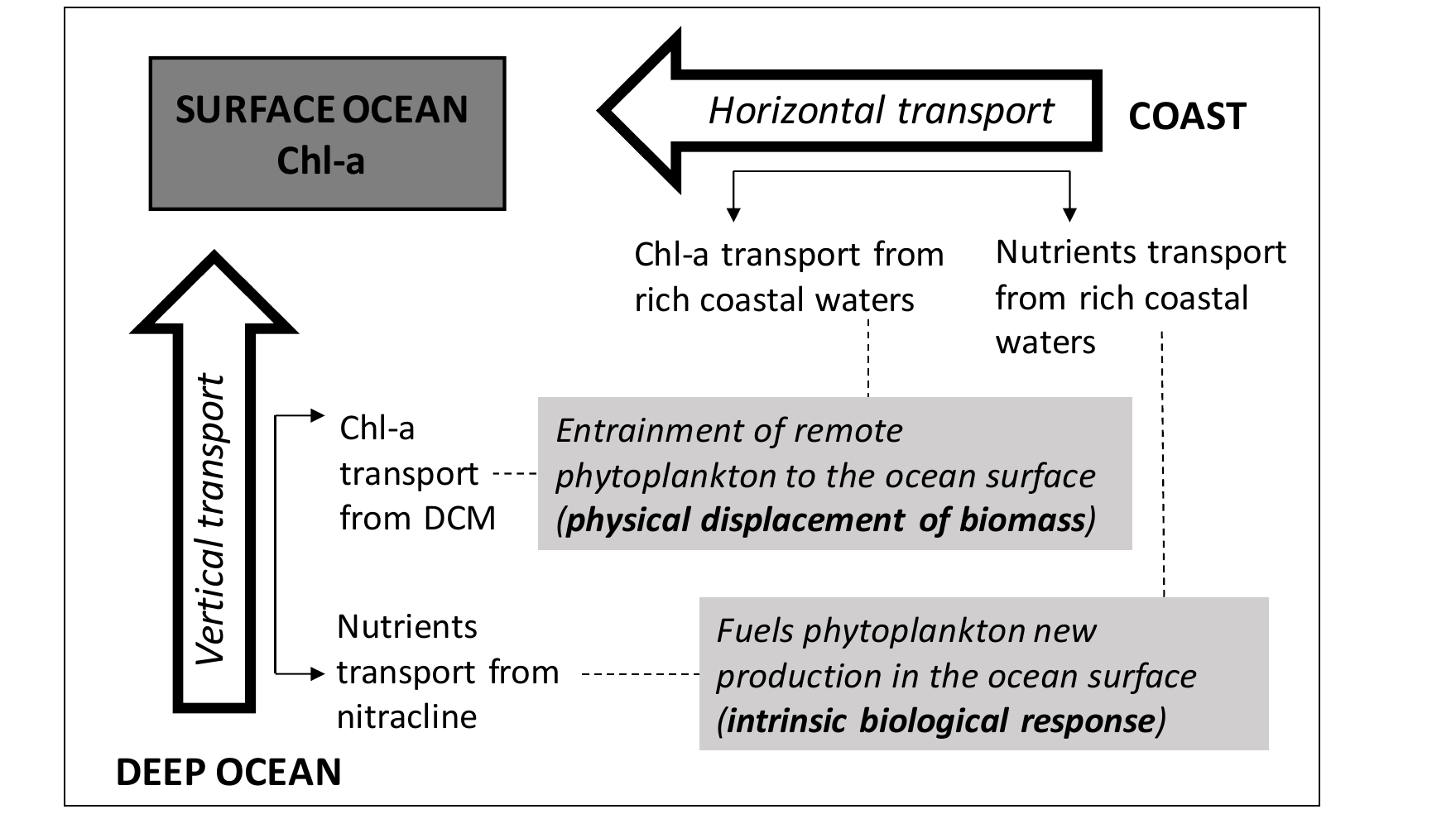}
\caption{Mechanisms by which a chlorophyll-a (chl-a) response can be detected following TC activity in ocean waters near coastal zones.}
\end{figure}

\end{document}